\documentclass[galaxies,article,accept,pdftex,moreauthors]{Definitions/mdpi} 
\firstpage{1} 
\pubvolume{1}
\issuenum{1}
\articlenumber{0}
\pubyear{2026}
\copyrightyear{2026}
\externaleditor{Firstname Lastname}
\datereceived{28 May 2024} 
\daterevised{22 January 2026} % Comment out if no revised date
\dateaccepted{ } 
\datepublished{ } 
\graphicspath{{./}{images/}}

\newcommand{\arcsec}{\mbox{$^{\prime\prime}$}}
\newcommand{\kms}{$\mathrm{km~s}^{-1}$}

\Title{IFU Spectroscopic Study of the Planetary Nebula Abell 30: Mapping the Ionisation and Kinematic Structure of the \\Inner Complex}

\Author{Kam %MDPI: Please carefully check the accuracy of names and affiliations.
 Ling Chan %MDPI: 1. We removed current address symbol and the note below as it is the same as affiliation 1. Please confirm. - okay 2. We removed invalid ORCID of this author, please confirm. - valid ORCID added
 $^{1,\dagger}$\orcidA{}, Andreas %MDPI: Order of author names is different from the system, please check and confirm. - author sequence now corrected according to authorship change form
 Ritter $^{1,}$* %MDPI: The corresponding author is different from the ones submitted online at susy.mdpi.com. Please confirm which are correct. - AR as corresponding author correct and according to authorship change form
$^{,\dagger}$\orcidB{}, Quentin Andrew Parker $^{1,\dagger}$\orcidC{} and Katrina Exter %MDPI: 1. This author is not listed in our system, please check and confirm or revise if necessary. - author sequence now correct according to authorship change form    2. Please add the email of this author to the affiliation 2 below. - done
 $^{2}$\orcidD{}}

\AuthorNames{Kam Ling Chan, Andreas Ritter, Quentin Andrew Parker and Katrina Exter}

\address{%
$^{1}$ \quad Laboratory for Space Research, Faculty of Science, The University of Hong Kong, Cyberport 4, \linebreak Hong Kong SAR, China; kamilachan@hotmail.com (K.L.C.); quentinp@hku.hk (Q.A.P.) %MDPI: We added these email addresses here according to those submitted online at susy.mdpi.com. Please confirm. - confirmed
\\
$^{2}$ \quad Institute of Astronomy, KU Leuven, Celestijnenlaan 200D, 3001 Leuven, Belgium; katrinaexter@gmail.com %MDPI: Please provide all information about this affiliation, including the institution names, postcode and city. - done
\\}

\corres{Correspondence: aritter@hku.hk %MDPI: The email address highlighted is different from the one submitted online at susy.mdpi.com. Please confirm which is correct. - okay
}

\firstnote{These authors contributed equally to this work.}  % Current address should not be the same as any items in the Affiliation section.
\abstract{This work presents integrated flux and velocity channel maps of the planetary nebula Abell~30 (A30) inner knot system. The observations were taken with the INTEGRAL spectrograph at the William Herschel Telescope (WHT), La Palma, Spain. Our IFU data cube has a field of view (FoV) of 12.3\arcsec $\times$ 16\arcsec~that partially covers knots J1 and J2, and completely covers knots J3 and J4 in the system. Optical Recombination Lines (ORLs) of C~II, He~I, He~II, N~III, O~II and Collisionally Excited Lines (CELs) of [Ar~IV], [Ar~V], [N~II], [Ne~III], [Ne~IV], and [O~III] were detected. Our integrated flux maps visualise the ionisation structure and the chemical inhomogeneity in the system previously reported by other groups. We find that ORLs are concentrated in the polar region (J1, J3), whereas the equatorial knots (J2, J4) are dominated by CELs. The flux ratio map of the diagnostic [O~III $\lambda$ 5007/4363 \AA] lines reveals the electron temperature distribution, which shows cold cores of 15,000 K in knots J3 and J4 surrounded by a hot outer layer of above 20,000 K. Our channel maps show positive and negative velocity excursions from the systemic value among the ions. Several ions show variation in their velocity structures from their lower-energy-level counterparts, including [Ar~IV] and [Ar~V], [Ne III] and [Ne~IV], and He~I and He~II. New recurrent velocity structures are identified in the low-density regions where the ions move much faster compared to their surrounding environments. %He~I, He~II, [O~III], [Ne~III], [Ne~IV], and C~II ions are found to be traveling in the same direction (red-shifted), while He~I 4026.08\AA\space and He~II 6560.10\AA\space progresses to another direction (blue-shifted).
The velocity dispersion measurements highlight extreme turbulence in some of the ions ($\sigma_{\mathrm{v_{rad}}} \approx 140$ km/s), consistent with supersonic/hypersonic motion driven by shocks. The forbidden line species [N~II] exhibits lower turbulence ($\sigma_{\mathrm{v_{rad}}} \approx$ 50--60 km/s), tracing denser, less-turbulent gases. Based on our data, we conclude that both the ionisation and kinematic studies hint at shock heating and multiple ejection history in the evolutionary pathway of A30.}

\keyword{(ISM:) planetary nebulae: individual; stars: kinematics; stars: mass-loss; 
(stars:) binaries: general; stars: late-type; stars: Wolf--Rayet} 

\begin{document}

\section{Introduction}

Abell 30 (A30) is a well-known and interesting planetary nebula (PN) that has a few 
different nomenclatures, e.g., PNG 208.5 + 33.2. It was known as PN A66 30 when it was first discovered in 1964 \cite{Greenstein1964,Abell1966}. 

A30 belongs to a highly ionised PN class arising from 
its very hot core. In the work of Jacoby \cite{Jacoby1979}, it was emphasised that 
A30 had a hydrogen-rich outer shell and a hydrogen-poor inner knot system 
in close proximity to the central star. This is rare in PNe
systems \cite{Jacoby1979}. A secondary envelope ejection was suspected 
as being responsible for the inner hydrogen-poor nebula \cite{Liu2000}.
Jacoby and Ford \cite{Jacoby1983} suggested 
the possibility of a binary companion which gained observational support by the detection of a variable light curve with a period of 1.060 days discovered in 2020 \cite{Jacoby2020}. The binary system in A30 is one of the proposed mechanisms to explain the chemical inhomogeneity observed in the inner knot system via the possible existence of an accretion disk from binary interaction \cite{Harrington1996}. Recent studies of the equatorial knots include refs.
\cite{Simpson2022,Toala2021}, which reveal the presence of dust in the equatorial region. 

In general, a main sequence star of mass ranging from 1 
to 8 solar masses will only eject its outer atmospheric layer once the object enters the AGB 
phase. The ejected material then becomes ionised by the hot core, causing it to glow in 
emission lines across much of the electromagnetic spectrum. This short 
evolutionary stage is collectively known as the PN phase. 
Its typical life span is about 5000--25,000 years \cite{Badenes2015}. PNe systems are generally 
hydrogen-rich \cite{Kwok2000}. Observational data for A30 suggests an interval of 12,000--18,000 years between 
the first and second ejection of the outer hydrogen-rich nebular shell and the inner hydrogen-deficient knot system \cite{Jacoby1979,Guerrero1996}, making it an {outlier relative to} %EE: please check intended meaning has been retained. - ok
the usual low-mass star evolutionary~pathway. 

The origin of chemically inhomogeneous ejecta in the inner knot system remains uncertain, and several theories have been proposed over the years. In the single-star scenario, a late-stage eruption via a Very Late Thermal Pulse (VLTP) \cite{Herwig2001,Toala2015} and subsequent born-again event \cite{Iben1983} explains the extreme abundance contrast between the H-rich outer nebula and the H-poor, C/O-rich knots surrounding the central star. Alternatively, the binary-star scenario involves multiple ejections modulated by a companion star \cite{Jacoby1983,Harrington1996}, including common envelope evolution \cite{Rodríguez2022}, to produce the observed clumpy, chemically distinct knots. The latest observational evidence clearly favours the binary scenario. 

\begin{figure}[h]
%\centering
\hspace{-8pt}\includegraphics[height=5cm]{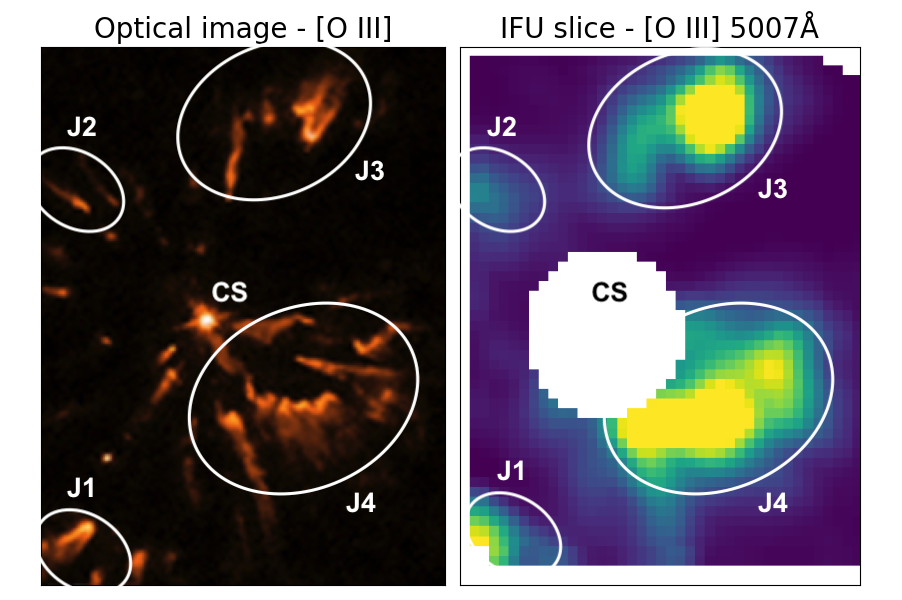}
\caption{Left panel: Gas clump identification of  knots J1--J4 from the outlined regions in white overlaid on the [O~III] optical image of the central inner region of A30, adopted from the Hubble Space Telescope WFPC2 \cite{Borkowski1995}. North is up and East is to the left.
Right panel: An integrated flux map of [O~III] 5007 \AA~as an example of the IFU sampled area with a FoV of 12.3\arcsec $\times$ 16\arcsec. Note that the central star (CS) and its surrounding region were masked.}
\label{fig:fig1}
\end{figure}

The hydrogen-poor inner knot system of A30 referred to in this work is divided into four main parts. These are knots J1, J2, J3, and J4 (Figure~\ref{fig:fig1}), starting from the SE nebula structure and counting clockwise around the central star \cite{Jacoby1979}. 
Knots J2 and J4 are collectively known as the equatorial system and knots J1 and J3 as the so-called polar knots.  

Here, we present an IFU spectroscopic study for A30 to investigate the distribution 
of ionic species across the gas clumps in the inner knot system and the kinematic structure 
observed in a series of spectral lines. Our integrated flux maps visualise the ionisation structure of detected ions that originates from various transitions taking place across the system. 
The channel maps and velocity structure visualised in our study support the multiple ejecta scenario or a complex outflow based on the varying Doppler shifts observed across different ions in the system.

\section{Observations}

Observations were taken using the INTEGRAL \cite{Arribas1998} integral field unit (IFU) 2D spectroscopic facility of the William Herschel Telescope in combination with the WYFFOS spectrograph. ``Fibre bundle 2'' with a FoV of 12.3\arcsec $\times$ 16\arcsec was used with an individual fibre size of 0.9\arcsec. Note that the 219 fibres are not spatially contiguous because approximately 40$\%$ of the total FoV falls into the gaps between the fibres. A ring of 30 fibres offset by 45\arcsec\ from the centre of the FoV allows the sky spectrum to be obtained simultaneously. The observations reported here were taken in 2004 (January 14) by Katrina Exter (PI) and were approximately centred on the central star of A30, fully covering the knots J4 and J3, and part of the equatorial knot J2 and the polar knot J1. Seeing was reported as $\le$1\arcsec. Six 1800 s observations with each of the red and blue 1200 l mm$^{-1}$ gratings were taken, covering the spectral range of 3800--5230 \AA~in the blue and 5170--6600 \AA~in the red. Our IFU data cubes are composites of 1001 slices with a wavelength spacing of 1.47 \AA~in the blue and red.

Bias subtraction, spectral identification, tracing, and extraction of the raw CCD data were performed within IRAF. Cosmic rays were removed with a median-combination technique (with $\sigma$-rejection) using images in groups of three, supplemented occasionally by manual intervention. The wavelength calibration was performed using spectra extracted from CuAr + CuNe arc calibration lamp frames. Due to the absence of arc emission lines below 4300 \AA~in the blue spectra, we re-calibrated those spectra using the measured positions of the PN emission lines reported in \cite{Wesson2003} for knots J1 and J3. %\cite{Wesson2003} do not state explicitly whether they corrected their spectra for the heliocentric radial velocity. 
We note that for the red spectra, our measured line positions show an offset of $\approx$1 \AA~compared to the positions stated in \cite{Wesson2003} (their wavelengths being smaller); however, the velocity maps from our wavelength calibration are more consistent between the blue and red arms for the individual line species (see below).

\section{Data Cubes}

Because all spectra were slightly offset with respect to each other, we converted the individual spectra (from 0.9\arcsec\ round spaxels with 40\%\ in gaps) into data cubes (with 0.3\arcsec\ spaxels with no gaps and a square grid) using the spatial interpolation routine provided by the ``Euro 3D visualisation tool'' \cite{Sanchez2004}. The core of this process is to resample the input spatial grid (which is the sky footprint of INTEGRAL) onto a new, contiguous, and regular grid, using ``Delaunay'' triangulation to select the best grid points. The fluxes are then interpolated from the input grid to the output grid at each wavelength in the dataset; we adopted the natural-neighbour flux-conserving interpolation method. The central star is a sufficiently bright source to trace the Differential Atmospheric Refraction (DAR) with wavelength, even though the PSF is not very well sampled \cite{Meneze2018}. After correcting for the DAR, the six red and blue cubes were combined into the final data cubes.

\section{Spectral Analysis---Integrated Flux Maps}

Abell 30 has been extensively studied for its asymmetric chemical distribution across the inner knot system, with various groups exploring the X-ray, UV, visible, IR, and FIR spectral ranges (e.g., \cite{Guerrero2012,Harrington1984,Cohen1977}). In the visible range, refs. \cite{Jacoby1983,Guerrero1996} and subsequent studies by refs. \cite{Simpson2022,Wesson2003} reported the presence of ORLs and CELs (or forbidden lines, demarcated by brackets []) of Oxygen (O~II, [O~II], and [O~III]), Neon ([Ne~III], [Ne~IV], and [Ne~V]), Nitrogen ([N~II] and N~III), Helium (He~I and He~II), Carbon (C~II and C~III), Sulfur ([S~II]), Argon ([Ar~IV], [Ar~V]), as well as trace amounts of Hydrogen ($\mathrm{H}_\alpha$ to $\mathrm{H}_\gamma$). Furthermore, ref. \cite{Toala2021} studied the carbon-rich dust and detected lines of heavier elements including Fe~V, Mg~I, and [Na~II] in the IR spectra, likely sourced from the equatorial region. Due to the presence of high-excitation emission lines, A30 is classified as a member of the high-excitation PNe.

Our spectral analysis focuses on the inner knot system within a 12.3\arcsec $\times$ 16\arcsec~FoV. Python 3.9 packages including PyNeb\endnote{\url{https://github.com/Morisset/PyNeb\_devel}, accessed on 15 January 2024}
 \citep{PyNeb2013} were used for the emission line analysis. The flux calibrated spectra of the individual spaxels were corrected for the heliocentric velocity and de-reddened using the extinction law of \cite{Cardelli1989}, adopting a value of c($\mathrm{H}_\beta$) = 0.6 from \cite{Wesson2003}, followed by continuum subtraction. To extract the ionisation structure information from our IFU data cube, the central star as well as the background star to the South--East of it were masked.

\begin{figure}[H]
%\centering
\includegraphics[width=37mm,height=45mm]{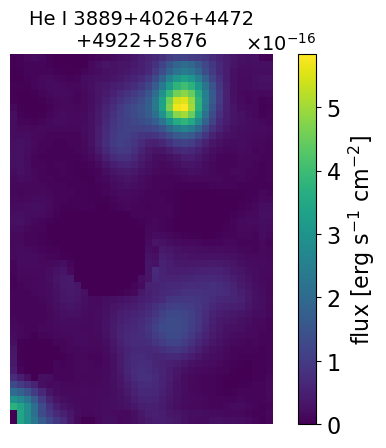}
\includegraphics[width=37mm,height=43mm]{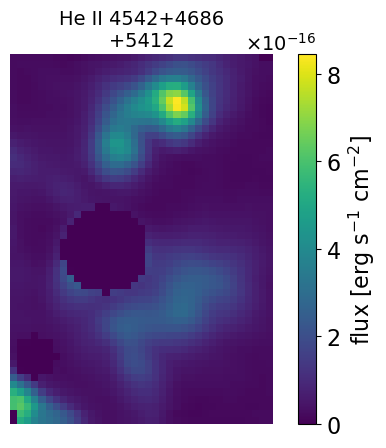}
\caption{Integrated flux maps of spectral lines of He~I (\textbf{left}) and He~II (\textbf{right}). Note the masked central star and background star.}
\label{fig:fluxHeI+II}
\end{figure}

\begin{figure}[H]
%\centering
\includegraphics[width=40mm,height=43mm]
{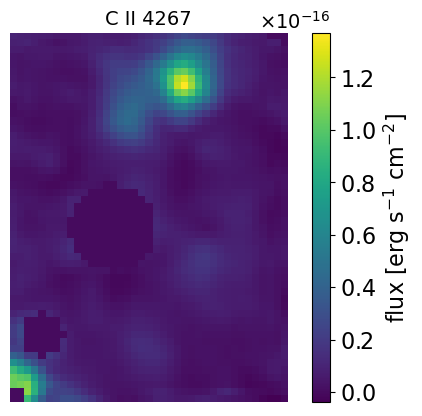}%Blue_final_cube_angstroms_2D_integrated_C II_4267.png}
\caption{The same as Figure~\ref{fig:fluxHeI+II}, but for C~II 4267 \AA.}
\label{fig:fluxCII}
\end{figure}

\begin{figure}[H]
%\centering
\includegraphics[width=40mm,height=43mm]{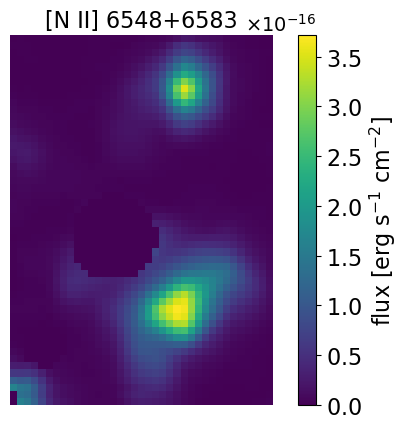}
\includegraphics[width=37mm,height=43mm]
{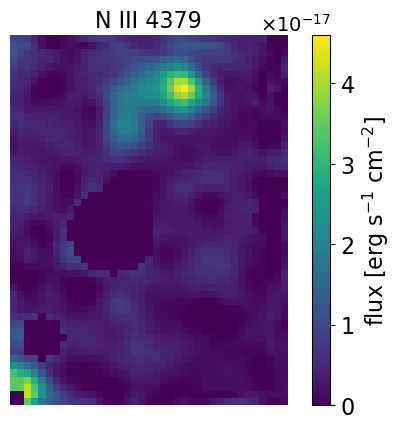}%Blue_final_cube_angstroms_2D_integrated_N III_4379.png}
\caption{The same as Figure~\ref{fig:fluxHeI+II} but for [N~II] 6548 \AA~+  6583 \AA~(\textbf{left}) and N~III 4379 \AA~(\textbf{right}).}
\label{fig:fluxNII+III}
\end{figure}

\begin{figure}[H]
\includegraphics[width=40.5mm,height=45mm]{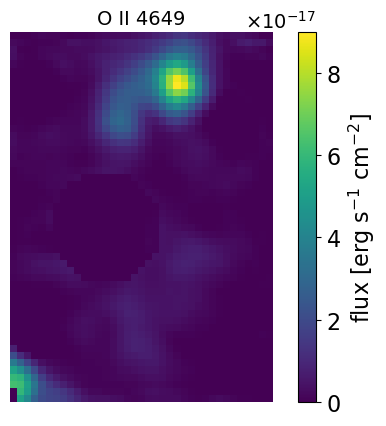}
\includegraphics[width=40.5mm,height=45mm]{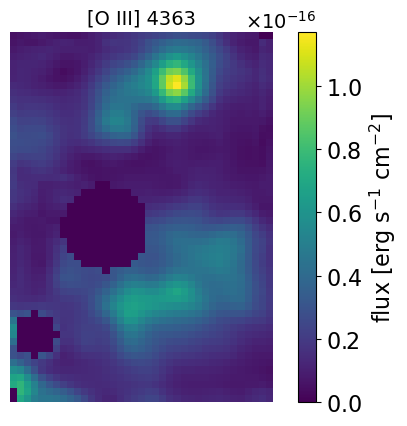}
\includegraphics[width=38.5mm,height=45mm]{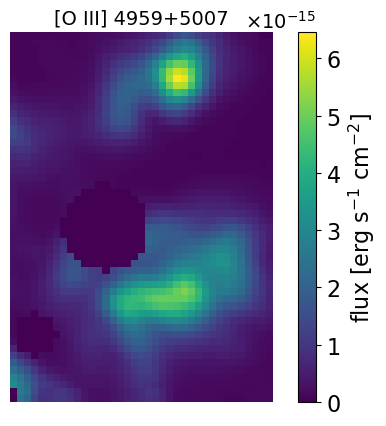}
\caption{The same as Figure~\ref{fig:fluxHeI+II} but for (from left to right) O~II 4649 \AA, [O~III] 4363 \AA, [O~III] 4959 \AA~+ [O~III] 5007 \AA. Note that the 4363 \AA\ line and the sum of 4959 \AA~and 5007 \AA\ are shown separately as they carry the electron temperature information (see below).}
\label{fig:fluxOII+III}
\end{figure}

\begin{figure}[H]
\includegraphics[width=37mm,height=43mm]
{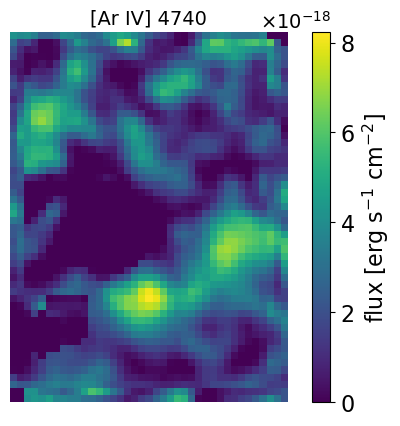}
\includegraphics[width=37mm,height=43mm]
{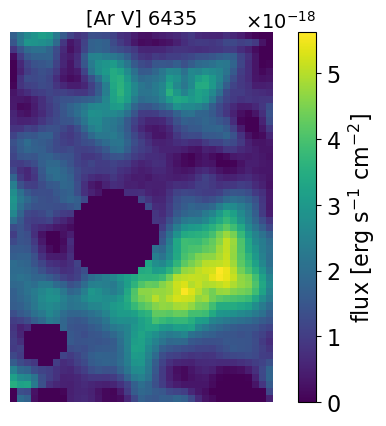}
\caption{The same as Figure~\ref{fig:fluxHeI+II} but for [Ar~IV] 4740 \AA~(\textbf{left}) and [Ar~V] 6435 \AA~(\textbf{right}).}
\label{fig:fluxArIV+V}
\end{figure}

\begin{figure}[H]
%\centering
\includegraphics[width=40mm,height=43mm]{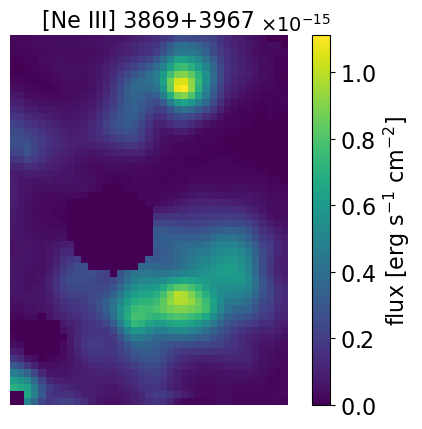}
\includegraphics[width=40mm,height=43mm]
{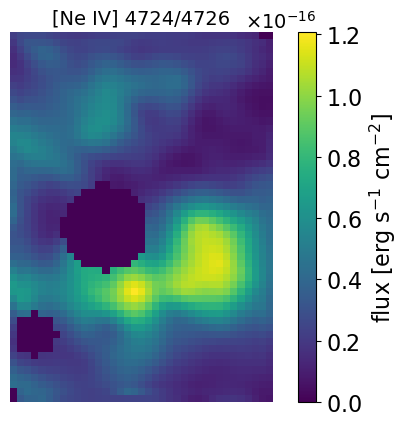}
\caption{The same as Figure~\ref{fig:fluxHeI+II} but for [Ne~III] 3869 \AA~+ 3967 \AA~(\textbf{left}) and [Ne~IV] doublet 4724/4726~\AA~(\textbf{right}).}
\label{fig:fluxNeIII+IV}
\end{figure}

For each un-blended spectral line detected, due to the skewed instrumental line profiles, we integrated the flux density over the corresponding wavelength range \mbox{($\lambda_0\pm6$ \AA)} to produce one single flux map. Individual fluxes of different lines from the same ion were then added to produce the final flux maps as shown in Figures~\ref{fig:fluxHeI+II}--\ref{fig:fluxNeIII+IV}. 
These maps visualise the ionisation structures of different ions distributed across the inner knot system that originate from various electron transitions. Table~\ref{tab:ionisation} shows the sum of fluxes for the individual ions and knots, scaled so that the strongest line ([O~III] in knot J4) has a strength of 10,000. {The ORLs of low-to-medium ionisation energy ions} %EE: please check intended meaning has been retained. - ok
mainly populate the polar knots J1 and J3 while the equatorial knots J2 and J4 are dominated by {the CELs of high-ionisation energy species}. %EE: please check intended meaning has been retained. - ok
The only recombination lines which can clearly be seen in the equatorial knot J4 are from Helium (He~I and II) atoms.

%Our flux maps show the ionisation structures of different ions distributed across the inner knot system that originate from various electron transitions. %\textcolor{red}{Findings from our maps mostly agree with  \cite{Jacoby1983} where He~I 3890\AA\Space, He~I 4471\AA\space and C~III 4647\AA\space are mostly found on knot J3 and faint in knot J4; absence of [Ar~IV] 4740 \AA\space line and \cite{Guerrero1996} where the [Ne~III] 3869 \AA\space is found among all knots, same for the C~II 4267\AA\space and C~III 4647\AA\space where it mostly concentrated in the polar knots...etc The electron temperature distribution was mapped based on the diagnostic [O~III] line ratio - I4363/(I4959 + I5007). }

%\textcolor{red}{Ions originating from the same upper energy level may display similar ionisation structures when exposed to the same radiation field \cite{Osterbrock2006}. Factors that explain the observed variation in ionisation structure despite the same upper energy level transition include (1) shock waves that lead to compression of the ionised gas, (2) ionisation environment - differences in pressure, temperature and density, (3) non-uniform radiation intensity which decides the rate of ionisation, (4) magnetic and electric fields that affect the motion of charged particles and (5) stellar winds from the central star \cite{DeMarco2009}.}

\begin{table}[H]
\caption{Ion species with ionisation energies needed to produce the ion and integrated fluxes of each species for each  normalised knot, so [O~III] in J4 is 10,000. Ions are sorted by their ionisation energies. Note that the different sizes of the knots influence the integrated fluxes, which may differ from the perceived strengths in the maps. Only J3 and J4 are fully covered by our FoV.\label{tab:ionisation}}
        \begin{tabularx}{\textwidth}{LCCCCC}
            \toprule
      \textbf{Ion} %MDPI: 1. We removed vertical lines, added bold for table header and formatted all tables according to layout rules, please confirm. - ok 2. Please confirm alignment in all tables. - ok
 & \textbf{Ionisation Energy [eV]} & \textbf{J1} & \textbf{J2} & \textbf{J3} & \textbf{J4} \\
            \midrule
            He~I & 0 & 89&19&391&233\\
            
            C~II & 11.3 & 27&4&95&32\\
            
            O~II & 13.6 & 17&0&59&14\\

            [N~II] & 14.5 & 30&11&152&396\\
            
            He~II & 24.6 & 219&96&793&812\\
            
            N~III & 29.6 & 11&4&33&13\\
            
            [O~III] & 35.1 & 882&783&4503&10,000\\
            
            [Ar~IV] & 40.7 & 1&4&5&15\\
            
            [Ne~III] & 41.0 & 156&144&784&1986\\
            
            [Ar~V] & 59.4 & 1&1&4&14\\
            
            [Ne~IV] & 63.5 & 20&57&71&337\\
            \bottomrule
        \end{tabularx}
\end{table}
\unskip
%\FloatBarrier % Place before or after the figure

%\FloatBarrier % Place before or after the figure
Neutral and singly ionised Helium lines mainly populate the polar knots J1 and J3, with traces to be found in J4 (Figure~\ref{fig:fluxHeI+II}). He~II lines appear slightly stronger in the South--East extension of J3 as well as in J4 compared to He~I. 
The ORL of C~II (4267 \AA, Figure~\ref{fig:fluxCII}, left) is only visible in the polar knots J1 and J3.

The forbidden singly ionised Nitrogen species ([N~II] 6548 \AA~and 6583 \AA, same upper and lower energy levels) covers knots J3 and J4 (see Figure~\ref{fig:fluxNII+III}, left and centre) and are fainter in J1. However, ORLs of N~III (4379 \AA, Figure~\ref{fig:fluxNII+III}, right) with twice the ionisation potential are only present in the polar knots J1 and J3. While [N~II] is concentrated in the North--West part of J3, N~III is also visible in the South--East extension of J3.

%Both spectral lines $\lambda$ 3869\AA~and $\lambda$ 3967\AA~of [Ne~III] (Figure~\ref{fig:fig2} left panel) exhibit complementary ionisation structure, consistent with their transition from the same level. The unresolved [Ne~IV] doublets $\lambda$ 4714/4716\AA~and $\lambda$ 4724/4726\AA~(Figure~\ref{fig:fluxNeIII+IV} right panel) are concentrated at regions covering the equatorial knots J2, part of the knot J4, and the region south west to the central star. 

%\FloatBarrier % Place before or after the figure

%The [N~II] spectral lines $\lambda$ 6548\AA~and 6584\AA~(Figure~\ref{fig:fluxNII+CII} left panel) show similar ionisation structure covering knots J1, J3 and J4. The ionisation maps of C~II ($\lambda$ 4267\AA) and C~III ($\lambda$ 4647\AA) exhibit similar structures primarily covering the polar knots (see Figure~\ref{fig:fig3} right panel) despite their different upper and lower levels where the transitions take place.
O~II (Figure~\ref{fig:fluxOII+III}, left panel) is also predominantly present in J1 and J3, while the forbidden [O~III] lines can also be found in J2 and J4 (Figure~\ref{fig:fluxOII+III} centre and right panels).
%\FloatBarrier % Place before or after the figure

%\FloatBarrier % Place before or after the figure

The high-ionisation species [Ar~IV] and [Ar~V] are very weak and noisy; however, it is clear that [Ar~IV] (Figure~\ref{fig:fluxArIV+V}, left panel) appears strongest in the equatorial knots J2 and J4, while [Ar~V] with an ionisation energy 1.5 times stronger appears to be only present in J4 (Figure~\ref{fig:fluxArIV+V}, right panel).

CELs of doubly ionised Neon ([Ne~III] 3869 \AA~and 3967 \AA, both representing transitions from the same upper level to the same lower level; see Figure~\ref{fig:fluxNeIII+IV}, left) are found both in the polar and equatorial knots. In contrast, its higher-ionisation counterpart [Ne~IV] (63.4 eV; unresolved doublet 4724/4726 \AA; see Figure~\ref{fig:fluxNeIII+IV}, right) that has a 1.6 times greater ionisation potential is predominantly found in the equatorial knots, especially knot J4. 

\begin{figure}[H]
%\centering
\hspace{-40pt}\includegraphics[width=60.5mm,height=45mm]{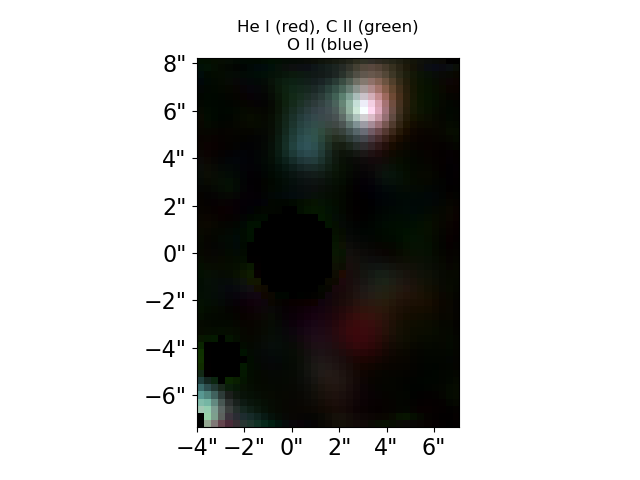}
\includegraphics[width=60.5mm,height=45mm]{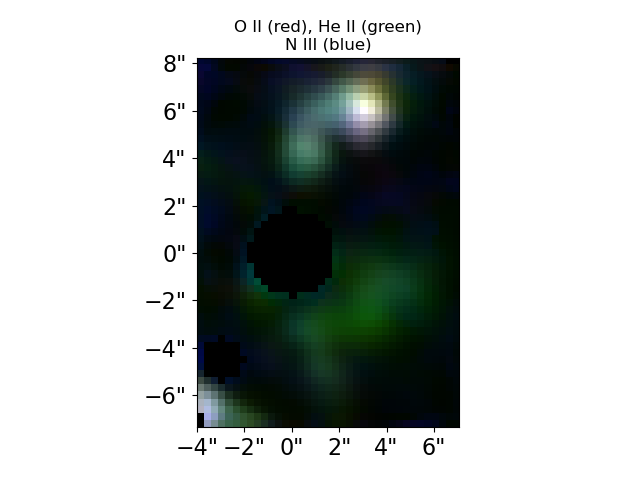}\\

\vspace{-12pt}
\hspace{-40pt}\includegraphics[width=60.5mm,height=45mm]{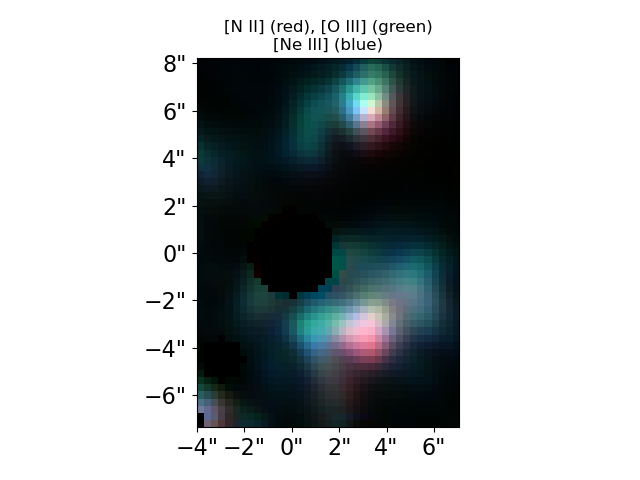}
\includegraphics[width=60.5mm,height=45mm]{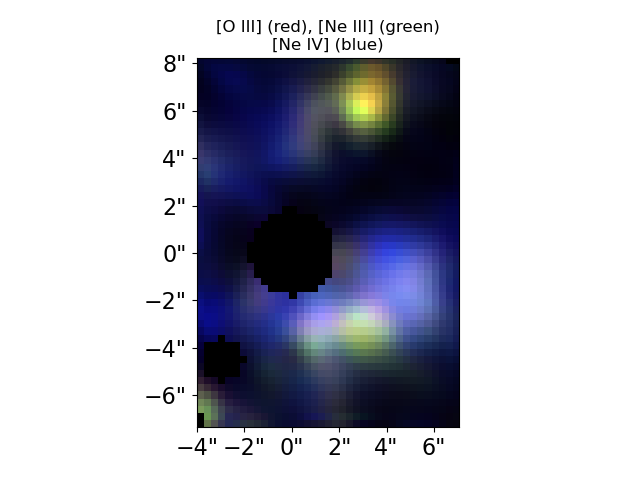}
\caption{Integrated flux maps of top: ORLs of left: 3 lowest ionisation species He~I (red), C~II (green), and O~II (blue); right: 3 highest ionisation species O~II (red), He II (green), and N~III (blue).
Bottom: CELs of left: 3 lowest ionisation species [N~II] (red), [O~III] (green), and [Ne~III] (blue); right: 3 highest ionisation species (except for Ar which is very weak and noisy) [O~III] (red), [Ne~III] (green), and [Ne~IV] (blue). Note that for each species multiple line fluxes were summed up and all fluxes were scaled to a maximum of 1.}
\label{fig:combinedFlux}
\end{figure}

Combinations of the 3 lowest and 3 highest ionisation species for ORLs and CELs individually are shown in (Figure~\ref{fig:combinedFlux}). Knot J4 is dominated by CELs with only a weak and diffuse contribution of He lines. Additionally, it appears that the lower-ionisation species are more concentrated compared to the higher-ionisation species, which are more diffuse. ORLs are produced predominantly in the polar knots J1 and J3 and partially in knot J4 of the equatorial knots. CELs of low- and high-ionisation-level species dominate the equatorial regions---J2 and J4.

\begin{figure}[H]
%\centering
\includegraphics[width=50mm,height=50mm]{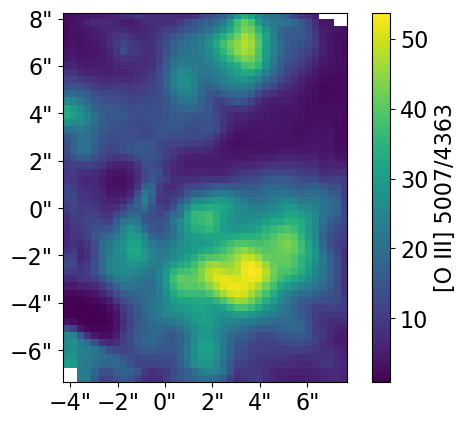}\hspace{10pt}
\includegraphics[width=55mm,height=50mm]{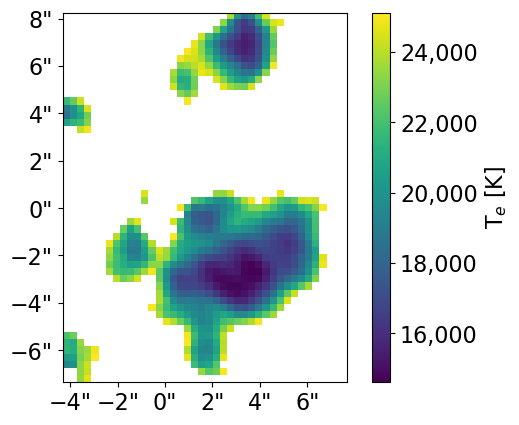}
\caption{(\textbf{Left}) [O~III] ratio map $\lambda$ 5007/4363 \AA. (\textbf{Right}) Electron temperature map derived from [O~III] ratio map on the left.}
\label{fig:temp}
\end{figure}

Our map of the electron temperature distribution (Figure~\ref{fig:temp}, right panel) was derived from the [O~III] line ratios 5007/4363 (Figure~\ref{fig:temp}, left panel). Ref. \cite{Wesson2003} state electron densities of 2800 and 3200 cm$^{-3}$ for the knots J1 and J3, respectively, which are the highest values obtained in the literature. However, even with these high values we are safely in the low density regime and assumed electron densities of 3000 cm$^{-3}$ for the creation of our temperature map. Knots J3 and J4 exhibit a cool core ($T_e\simeq15,000$ K) surrounded by a hot outer layer of above 20,000 K, which is in good agreement with previously determined averaged temperatures by \cite{Jacoby1983,Guerrero1996,Simpson2022} as well as the proposed existence of cold cores surrounded by much hotter outer regions of the knots \cite{Harrington1984, Wesson2003}. Knots J1 and J2 are only partially covered so no statement about a possible cool core and hot shell can be made here. The presence of cool cores may indicate shielding from the central star's radiation, possibly by dust \cite{Harrington1984}, or shock heating in the outer envelopes of the high-density knots.

%\textcolor{red}{The observed distribution suggests sources of high energy ionisation in the equatorial region. Examples for these high energy sources include shock heating or multiple ejection events that were subjected to different ionising conditions that gave rise to the high and low ionisation species observed in the equatorial and polar knots regions. In particular, regions West to the central star that cover part of knot J4 possess both high and low ionising species that hint at a complex, mixed ionising source in the region \cite{Osterbrock2006}.}
\vspace{-6pt}
%\FloatBarrier % Place before or after the figure

\section{Kinematic Structure and Velocity---Multiple Kinematic Components}
The outer shell of A30 expands at a velocity of about 30--40 \kms~while the inner knots show complex structures with velocity spikes of $\pm$200 \kms~\cite{Meaburn1996,Fang2014}. 

The spectral resolution of our data ranges from R $\approx$ 1650 at 3800 \AA~to R $\approx$ 2700 at 5200 \AA~for the blue arm and R $\approx$ 2100 to 3600 for the blue and red ends of the red arm, respectively. Doublets that are less than 3 \AA\space apart are therefore not resolved in our data, as is the case for [Ne IV] 4725.5/4726.9 \AA, leading to broadened line profiles. The spectral resolution is equivalent to a velocity resolution of $\approx$180~\kms~at the blue end and $\approx$80~\kms~at the red end. Discrete sampling intervals (sampling widths) of each slice are $\approx$115 \kms for the blue end and $\approx$70~\kms~for the red end, which give sample ratios of $\approx$1.57 to 1.14, respectively, which is lower than the Nyquist-compliant sampling requirement of $\geq$2 \cite{Nyquist1928}. Therefore, velocity analysis in the kinematic studies of this work is focusing on kinematic structure. All velocities derived in this study are based on the laboratory wavelength. 

\begin{figure}[H]
%\centering
\includegraphics[width=\textwidth]{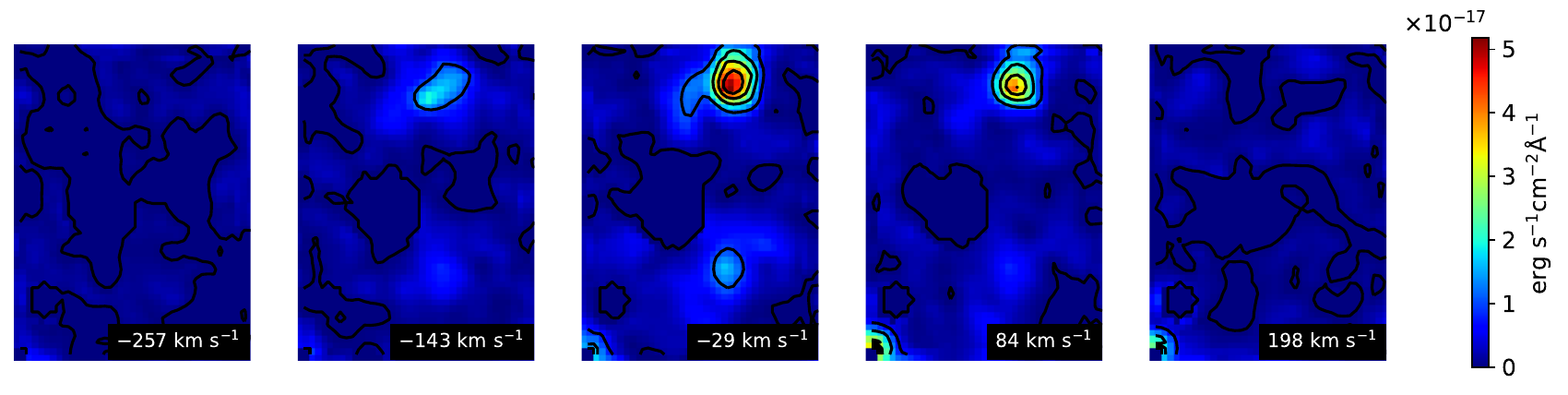}\\
\includegraphics[width=\textwidth]{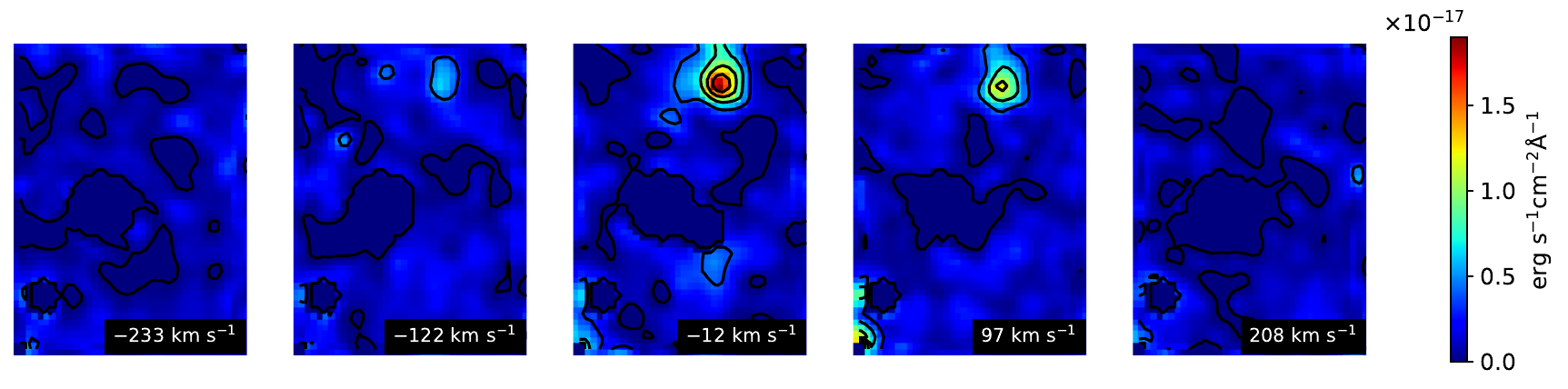}\\
\includegraphics[width=\textwidth]{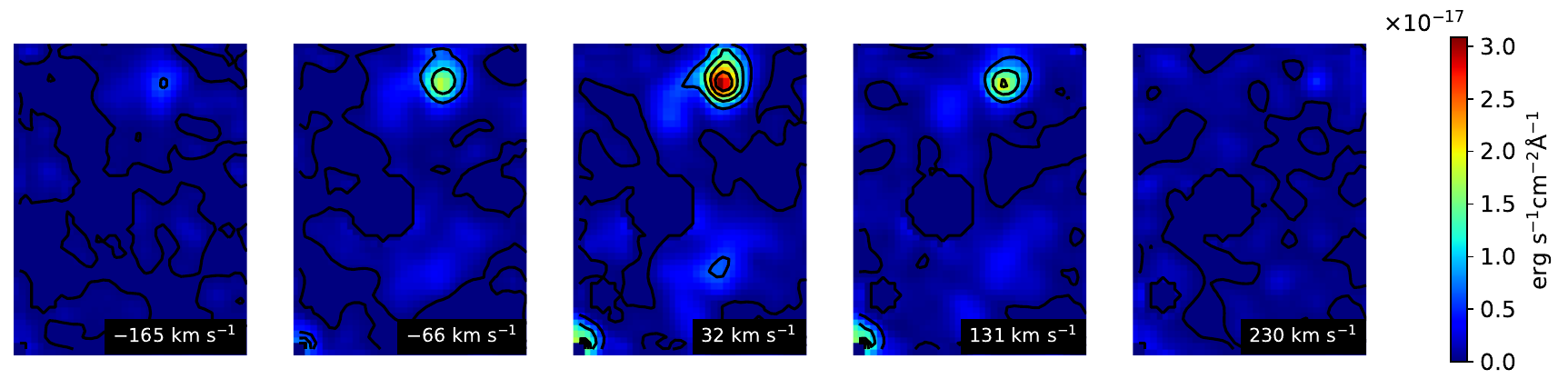}\\
\includegraphics[width=\textwidth]{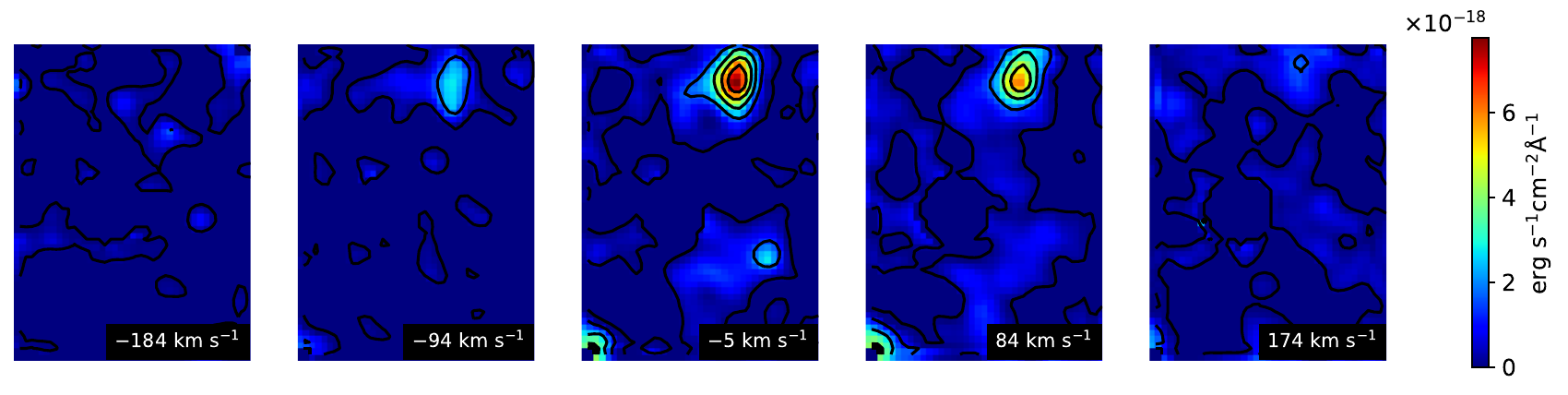}\hfill\\
\includegraphics[width=\textwidth]{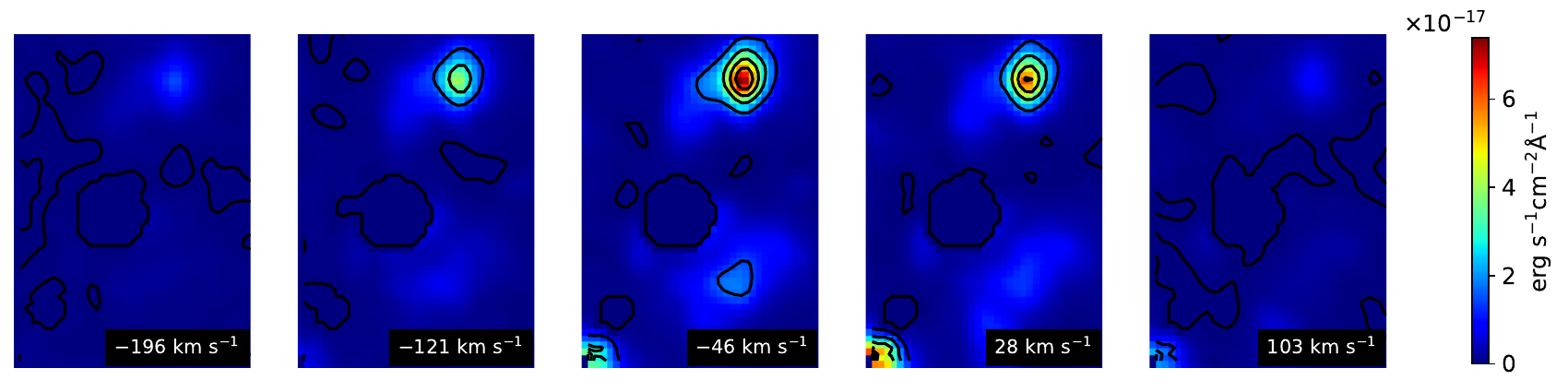}\\
\centerline{
\includegraphics[width=55mm,height=45mm]{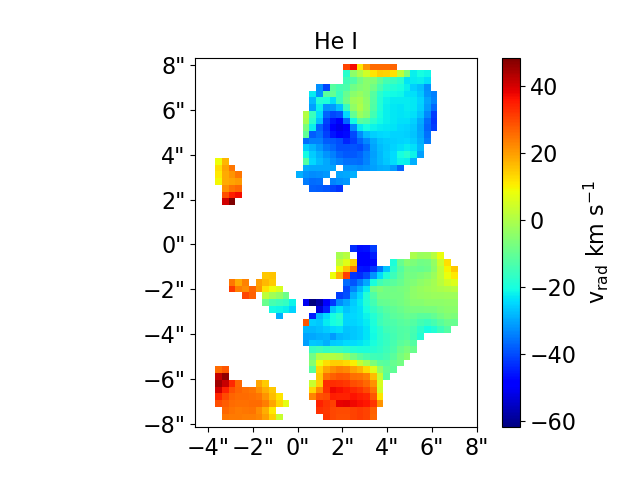}
\includegraphics[width=55mm,height=45mm]{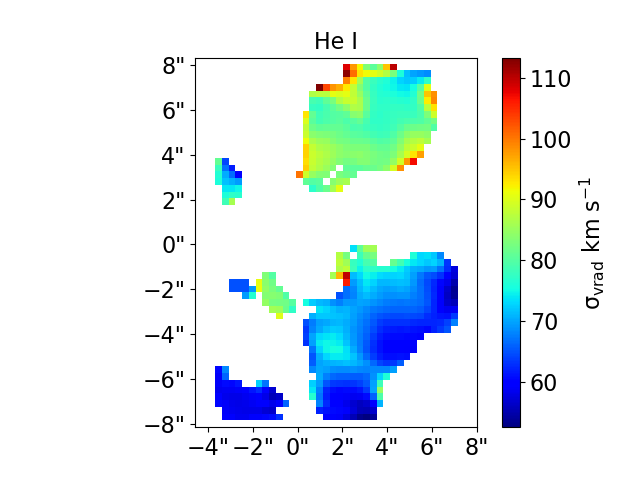}
}
\caption{Kinematic structure of He~I (from top to bottom) 3889 \AA, 4026 \AA, 4472 \AA, 4922 \AA, and 5876~\AA. The first 5 rows show the flux density values of the individual spaxels at velocities calculated with respect to the remaining wavelengths of the individual lines. The bottom row shows the mean velocities for each spaxel (left) as well as the velocity dispersions (in \kms, right).}
\label{fig:velHeI}
\end{figure}

We produced kinematic maps (Figures~\ref{fig:velHeI} and \ref{fig:velHeII}--\ref{fig:velNeIV} top panels) for each identified, un-blended line by plotting the measured flux density for each wavelength step and spaxel. %Line profiles were produced by selecting all emission lines covering a knot with at least 50\% of the flux level of the strongest emission line, normalising the lines by the maximum of a fitted Gaussian (with the outer wings and any negative values forced to 0), and plotting the mean value of each line for each wavelength/velocity step relative to their laboratory wavelength, using a 3$\sigma$ rejection algorithm. If there was more than 1 emission line of the same species, all lines were combined in an individual plot for each knot, allowing for a better representation of the line profile as each line is sampled at different wavelengths relative to its centre (Figures~\ref{fig:velNeIII}--\ref{fig:velOIII} center panels).\\
\textls[-15]{To produce the velocity maps at the bottom panels of  Figures~\ref{fig:velHeI} and \ref{fig:velHeII}--\ref{fig:velNeIV}, we fitted all lines of each ion species with signal-to-noise ratios (SNRs) greater than 3 simultaneously with the same velocity. As the velocity resolution changes with the wavelength, we set lower limits on the velocity dispersion of each line corresponding to the velocity resolution at those lines. The fitted mean velocities (relative to the laboratory wavelength) are shown in the left-hand plots of the bottom panels of each figure, while the velocity dispersions (in \kms) of the line with the highest velocity resolution (which is the reddest line) are shown on the right-hand side. %On the right-hand side of those plots we created different line-profile plots taking advantage of the fact that in case of different mean velocities the line profile will be sampled differently at a given wavelength grid. For this, each spaxel inside an individual knot was again fitted with a Gaussian, but this time we plotted the resulting line profile centered on the mean of each Gaussian fit. If the mean values (velocities) do not change then all plotted profiles share the same x values. If the profile changes but the mean value stays the same then there appear vertical lines for each wavelength step. Two such examples are the line profiles for the knot J1 and the [N~II] 6548\AA\, and 6583\AA\, lines (Figure~\ref{fig:velNII} bottom panels). If the mean value changes but the shape of the profile stays the same then we get the line profiles sampled at different positions, producing more continuous line profile plots, e.g. for J4 in Figure~\ref{fig:velNII}. If both the mean values and the profiles change then the resulting plot is more chaotic (see Figure~\ref{fig:velNIII} J3 for an example).
All the contoured kinematic maps presented here are in linear scale. The mean radial velocities of a Gaussian fit for each ion species summed up for each knot are shown in Table~\ref{tab:vrad}. We adopted the velocity uncertainties from the covariance matrix of the weighted least-squares fitting algorithm as error estimates. Note that the velocity dispersions stated in the table can arise either from actual turbulence or from differing velocities for the individual spaxels inside a knot. Additionally, the instrumental profile is wider in the blue region, leading to larger lower limits of the measured velocity dispersions compared to the lines in the red region.}

\begin{table}[H]
\caption{Mean radial velocities of Gaussian fits for each ion species and knot.\label{tab:vrad}}
\begin{tabularx}{\textwidth}{LCCCC}
\toprule
     & \multicolumn{4}{c}{\boldmath{$\mu_{\mathrm{vrad}}$} \boldmath{[\kms]}} \\
     %& \multicolumn{4}{c}{$\sigma_{\mathrm{vrad}}$ [\kms]}\\
\midrule
    \textbf{Ion} & \textbf{J1} & \textbf{J2} & \textbf{J3} & \textbf{J4} \\
    \midrule
    He~I & 31  $\pm$  21 & 56  $\pm$  33 & $-$25  $\pm$  22 & $-$17  $\pm$  26 \\
      He~II & 38 $\pm$ 18 & 48 $\pm$ 21 & $-$2 $\pm$ 18 & 16 $\pm$ 19 \\
       C~II & 62 $\pm$ 32 & 38 $\pm$ 68 & $-$2 $\pm$ 34 & 2 $\pm$ 83 \\
    %J2 questionable
    {[N~II]} & 9 $\pm$ 6 & 23 $\pm$ 14 & $-$26 $\pm$ 5 & $-$19 $\pm$ 4 \\
     N~III & 41 $\pm$ 26 & $-$68 $\pm$ 75 & $-$14 $\pm$ 23 & 25 $\pm$ 86 \\
    %J2 questionable      
    O~II & 56 $\pm$ 37 & & 12 $\pm$ 42 & 64 $\pm$ 93 \\    
    {[O~III]} & 32 $\pm$ 17 & 68 $\pm$ 17 & 11 $\pm$ 17 & 17 $\pm$ 17 \\    
   {[Ar~IV]} & 51 $\pm$ 98 & 53 $\pm$ 68 & $-$7 $\pm$ 79 & 16 $\pm$ 74 \\    
    {[Ar~V]} & $-$15 $\pm$ 41 & 51 $\pm$ 49 & $-$18 $\pm$ 41 & $-$27 $\pm$ 27 \\
    {[Ne~III]} & 55 $\pm$ 23 & 25 $\pm$ 22 & $-$11 $\pm$ 21 & $-$30 $\pm$ 21 \\    
    {[Ne~IV]} & 10 $\pm$ 37 & 44 $\pm$ 27 & $-$7 $\pm$ 38 & $-$11 $\pm$ 26 \\
    \midrule
    mean & 34 & 34 & $-$8 & 3 \\    
    stddev & 23 & 36 & 12 & 27 \\
    \bottomrule
\end{tabularx}
\end{table}

The He~I lines (see Figure~\ref{fig:velHeI}) are quite weak but still very clearly detectable, mainly in knots J1 and J3 and less so in J4. For J2, barely any detection was possible. He~I in J1 appears red-shifted by $\approx$30 \kms~while J3 and J4 are centred around $-$20 \kms. The velocity dispersions in J3 are greater compared to J1 and J4, indicating higher turbulence or multiple kinematic components in the line of sight.

The He~II lines (see Figure~\ref{fig:velHeII}) at 4542 \AA~and 5511 \AA~are very weak and subsequently quite noisy, while the 4686 \AA~line is reasonably strong and can be fitted across most of the FoV. The flux maps for the individual velocity steps of the 3 lines appear consistent. Knots J1, J2, and J4 appear to be slightly red-shifted, with the highest being J2 at $\approx$50 \kms, while J3 appears static. The lowest degree of turbulence is seen in J4, where the velocity dispersions are close to the lower limit set by the spectroscopic resolution. J2, however, shows velocity dispersions about twice as high. We note that the sharp edges of J4 in the velocity\hfill dispersion\hfill map\hfill are\hfill caused\hfill by\hfill the\hfill higher\hfill velocity\hfill resolution\hfill of\hfill the\hfill 5412 \AA~line
%\FloatBarrier % Place before or after the figure
\hspace*{-7.5mm}compared to the strongest line at 4686 \AA. As the 5412 \AA~line has a SNR below 3 in the areas surrounding J4, it was not included in the fits, explaining the sharp edges.

The C~II 4267 \AA~line (Figure~\ref{fig:velCII}) is only detected in our maps in knots J1 and J3. While J1 is strongly red-shifted with velocities close to 60 \kms, J3 shows velocities around $0\pm20$ \kms. J1 shows very little turbulence. The highest velocity dispersions with values exceeding 110 \kms~are seen between the higher-density sub-structures in J4.

\textls[-15]{The [N~II] lines at 6548 \AA~and 6583 \AA~(Figure~\ref{fig:velNII}) are sampled at similar velocities. Maximum velocities with values of $\approx$30 \kms~are reached in J2, South--East of J4, as well as just East of the CS, and slightly less in J1. J3 is slightly blue-shifted with velocities of \mbox{$\approx$$-$20 \kms}~while the highest negative velocities with values of $<$$-$40~\kms~are found South and West of the CS. J4 is the least turbulent, with velocity dispersions of the Gaussian fits of $\approx$45 \kms,~close to the lower limit set by the velocity resolution. J1, J2, J3 as well as the area South--East of J4 show more turbulent gas with velocity dispersions of $\approx$55--60 \kms. The highest turbulence---although the signal is quite faint---is found just South--East of the CS.}

The weak N~III line at 4379 \AA~(Figure~\ref{fig:velNIII}) is clearly red-shifted at J1 with mean velocities of $\approx$40 \kms~and mildly blue-shifted at J3. The velocity dispersions between 120 and \mbox{140 \kms}~are more than twice as high compared to the [N~II] lines, although influenced by the wider instrumental profile. The turbulence in J3 shows a systematic trend to increase towards North.

The only O~II line which is free of blends in our data is the 4649 \AA~line (Figure~\ref{fig:velOII}). The line is red-shifted for knots J1, J3, and J4, and not detected in J2. The highest velocities with values of up to $\approx$100 \kms are found in the southern part of J4. Turbulence is moderate with an apparent sea-saw structure in knot J4 (values between 60 \kms and 140 \kms) and relatively constant at $\approx$100 \kms in J1 and J3.

The CELs of [O~III] at 4363 \AA, 4959 \AA, and 5007 \AA~(Figure~\ref{fig:velOIII}) show the highest velocities at $\approx$100 \kms~West for J2 and between J3 and J4. The knots themselves show moderate red-shifts with velocities of $\approx$30--50 \kms~in J1, $\approx$60--70 \kms~in J2, $\approx$10 \kms~in J3, and $\approx$20--30 \kms~in J4. Turbulence is relatively low with values around 70--80 \kms~in knots J3 and J4, $\approx$90 \kms~in J2, and 90--100 \kms~in J1. The strongest turbulence with values of 100-120 \kms~is found East and North of the CS as well as between J3 and J4, with the highest values of up to 140 \kms~West of J3.

The emission lines of [Ar~IV] (4740 \AA, Figure~\ref{fig:velArIV}) and [Ar~V] (6435 \AA, Figure~\ref{fig:velArV}) are weak; however, some systematic movements can be seen. 
J1 and J2 for [Ar~IV] 4740 \AA~are red-shifted with strongly varying mean velocities between 0 and $+100$ \kms, with the highest velocities actually between the knots. Velocities of J3 and J4 on the other hand are centred around $\approx$0 \kms. The strongest turbulence with velocity dispersions exceeding 140 \kms~is found just South of J3, while the knots show values of 60--130 \kms. %Given the very low SNR not much can be said about the line profiles of the individual knots. 
[Ar~V] 6435 \AA~exhibits different kinematics compared to [Ar~IV]---while J2 is still strongly red-shifted with velocities up to $\approx$75~\kms, the other knots now appear blue-shifted with velocities between 0 and $-60$~\kms. J4 shows moderate turbulence of 50--100 \kms.%Changes in the mean values and line profiles are evident despite the low SNR.\\

%The kinematic morphologies of the He~I lines 3889\AA, 4026\AA, 4471\AA, 4922\AA, and 5876\AA\space(see Figure~\ref{fig:velHeI}) are similar, with most atoms being present in the polar knots J1 and J3. The He~I lines peak at velocities around 0 km/s. Both measured He~II lines (Figure~\ref{fig:fig9}) $\lambda$ 4865.68\AA\space and 6560.10 \AA\space are more prominent in knot 4 compared to He~I lines, moving with peak velocities of -69 \kms to -94 km/s respectively.

%However, not all of the He~I and He~II lines observed were consistent in their ionisation nor kinematic structure - i.e. He~II 4101\AA\space  possess a similar kinematic structure to He~I while that of He~I 5784\AA\space  is similar to He~II, despite transitioning from the same ionisation energy level. Such variation might have been given rise by (1) overlapping, (2) contamination by nearby spectral lines, (3) instrumental effect like low resolution and bad Point spread function (PSF) sampling. Therefore, the channel maps nor velocity calculation for those lines were not presented in this report.  

The kinematic morphologies of [Ne~III] lines at 3869 \AA\space and 3967 \AA\space(see Figure~\ref{fig:velNeIII}) indicate that J1 is quite strong, and J2 and J3 are slightly red-shifted, while J4 is slightly blue-shifted. The fastest moving ions---moving at speeds of $\approx$100~\kms---can be found in J1 and North--West of J4. Knots J2--J4 exhibit moderate turbulence with velocity dispersions between 90 and 110 \kms,~while J1 and the areas around the knots show larger values between 120 and 140 \kms.

\textls[-15]{The unresolved [Ne~IV] 4724/4726 \AA~doublet (assumed remaining wavelength 4724.89 \AA, Figure~\ref{fig:velNeIV}) shows that J2 is the most red-shifted with an average velocity of $\approx$50 \kms, J1 is slightly red-shifted, J3 shows velocities centred around 0 \kms with strong red-shifts just South of it, and J4 is slightly blue-shifted. All knots show velocity dispersions around \mbox{100--110 \kms~}while the area South of J3 shows values of up to 165 \kms.
%The fastest ions, moving at velocities of up to 100~\kms, are found just South and West of knots J2 and J3. 
%The line profiles centered on the mean velocity show strong variations in knots J1 and J2, mild variations in J3, and very little variations in J4.
The direct comparison between the [Ne~III] lines and the [Ne~IV] 4724/4726 \AA~doublet shows various differences. The mean velocities in the knots differ quite strongly with J1, showing a much lower velocity for [Ne~IV] ($\approx$20 \kms) compared to [Ne~III] ($\approx$70 \kms). In contrast, ions just South of J2 and J3 show maximum velocities of $\approx$100~\kms~for [Ne~IV], while both [Ne~III] lines only show velocities of $<$50~\kms.}

\section{Discussion}
\subsection{Multiple Kinematic Components---Non-Uniform Velocity Distributions}
\textls[-15]{Based on the work on A30 for the inner knot kinematic structure by \cite{Jacoby1989}, the knots possess different velocity components. The better resolved studies by \cite{Fang2014,Meaburn1996} also reached similar findings, suggesting that the knots were not generated from a single episode of matter outflow but several---a rather complex process. Velocities derived in our study show a non-uniform velocity distribution among the knots, e.g., different mean velocities for emission lines originating from different ions of the same elements. While some of them may simply be due to low SNR, others appear to be more significant. Still, even these more significant systematics may have been caused by different effects. It is possible that a Gaussian is not an appropriate representation of the emission lines, leading to wrong fitted mean values. Other possible  reasons are unidentified blends with lines from other atoms, data-reduction or wavelength-calibration problems/artifacts, or pixel-to-pixel correlations introduced by data (re-)binning. While our data show some evidence for this unexpected behaviour being real (supporting previous studies), only follow-up observations with higher spectral resolutions and higher SNRs can reveal the full picture.}

The highest velocities are found between the knots, likely tracing low-density material, which has been accelerated by the stellar wind, while the higher-density material is shielded from the stellar wind, possibly by dust grains. Larger velocity dispersions of the Gaussian fits in these low-density regions fit this picture and can be explained with turbulence, while large velocity dispersions in the high-density regions may be due to either turbulence or multiple kinematic components in the same line of sight.

The knots themselves show multiple different velocity components with varying mean values and velocity dispersions. While knots J1 and J2 are consistently red-shifted (except for [Ar~V] where J1 appears to be blue-shifted), J3 and J4 show red-shifted emission lines as well as blue-shifted ones. Notable outliers from the general velocity structures are the [N~II] lines at 6548 \AA~and 6583 \AA, which have the lowest velocity dispersions, and the [Ar~V] line at 6435 \AA, which is the only blue-shifted line in J1, red-shifted in J2, and the strongest blue-shifted line in J3 and J4.

As for the mean velocities for the individual knots stated in Table~\ref{tab:vrad}, it seems probable that the individual sub-knots visible in the HST image (Figure~\ref{fig:fig1}) have differing velocities, leading to several components within each integrated J complex. Follow-up observations with better spatial and velocity resolution are required to extend this study.

\textls[-15]{A new recurrent structure in the velocity maps, apart from the knots J1--4, is the circular area at 0\arcsec<x<3\arcsec, y<$-$5\arcsec, which, while not containing much flux itself, is consistently moving at mean velocities $\approx 50$ \kms~higher compared to the surrounding area. The same structure can be identified in Figures~\ref{fig:velHeI}, \ref{fig:velHeII}, \ref{fig:velNII}, \ref{fig:velArIV}, \ref{fig:velOIII}, \ref{fig:velNeIII}, and~\ref{fig:velNeIV}. Other new velocity structures moving with much higher velocities compared to their surrounding areas are the ones surrounding J2 and the area between J3 and J4, which are most prominent in Figures~\ref{fig:velOIII} and~\ref{fig:velNII}, but also observable in Figures~\ref{fig:velHeI},~\ref{fig:velHeII},~\ref{fig:velArIV},~\ref{fig:velNeIII}, and~\ref{fig:velNeIV}. A small area just East of the CS at y values between 0\arcsec~and $-$2\arcsec~appears strongly red-shifted for He~II while strongly blue-shifted for [O~III] and [Ne~III].}

\subsection{A30 and Its Massive Wolf--Rayet Counterparts}
A30 hosts a central star of the Wolf--Rayet spectral type that has been categorised as a stage in transition between the spectral-type Wolf--Rayet star and PG 1159: [WC]-PG 1159 \cite{Werner2024}. A number of massive Wolf--Rayet stars have been identified as either periodic or episodic dust makers \cite{Williams2019}. It is believed that the wind collision region created by binary interaction often gives rise to strong compression and subsequent radiative cooling that leads to dust formation in these systems \cite{Hendrix2016}. Examples of dust-making binary massive Wolf--Rayet systems that were detected with X-ray and radio emission include WR~140, WR~137, and WR~125 \cite{Lau2022,Lau2024,Richardson2024}, where binary-driven shock-triggered dust formation was observed in both WR~140 and WR~137 \cite{Lau2022,Zhekov2015}. Despite the long orbital period and periodic dust formation process of these Wolf--Rayet binary examples, physical phenomena, including dust formation, X-ray emission, and binary interaction, observed in these binary Wolf--Rayet cases make them a possible analogue for a better understanding of our A30 inner knot system characterised by its optical and X-ray signatures and suggested binary central star system.

%%%%%%%%%%%%%%%%%%%%%%%%%%%%%%%%%%%%%%%%%%
\section{Conclusions} 

We obtained integral field spectra of Abell 30's inner knot system with a FoV of 12.3\arcsec $\times$ 16\arcsec. After data reduction, we integrated the IFU data cube slices into emission line flux maps for the detected spectral lines of He I and II, C II, [Ne~III] and [Ne~IV], [Ar~IV] and [Ar~V], O~II and [O~III], [N~II] and N~III. 

ORLs of low-ionisation species predominantly populate the polar knots J1 and J3 while the CELs of high-ionisation species dominate the equatorial knots. Helium emission is strongest in J1 and J3, weaker in J4, and almost absent in J2. C~II is strong in J1 and J3 and barely detectable in J2 and J4. [N~II] is present in J1 and J3, and strongest in J4, where N~III is barely visible. O~II is strong in J1 and J3, absent in J2, and weak in J4, where strong but diffuse [O~III] is seen. The CELs of Argon are strongest in J4 and only barely detected in the other knots. [Ne~III] is present in all 4 knots while [Ne~IV] is strong only in J4 with diffuse emission across most of the FoV.
%The coexistence of ions of distinctive ionisation energy (from the lowest - C~II: 11.3 eV to the highest - [Ne~IV]: 63.5 eV) in J4 implies a mixed ionisation condition, possibly due to an overlapping shock and photo-ionisation zone.% The presence of high ionisation species [Ne~IV], [Ar~IV and V] in the equatorial region supports the hypothesis of shock heating history in the region that subsequently led to gas compression followed by formation of the Carbon-rich dust as previously reported by other groups.

The constructed temperature map based on the detected [O~III] diagnostic lines shows cool cores ($\approx$15,000 K) in the knots surrounded by hot outer regions exceeding 20,000 K, indicating possible shocks, shielding of the inner cores, or a combination of both.

For the velocity maps, we extracted IFU slices covering the detected spectral lines. %Velocities were calculated relative to the laboratory wavelengths of the spectral lines. 
Most observed lines in our data are red-shifted, with the strong exception of blue-shifted [Ar~V] lines in J1, J3, and J4 (while J2, even though the signal is very weak, is red-shifted). J1 and J2 (where detected) are consistently red-shifted (with the exception of [Ar~V] in J1) for all detected emission lines. J3 as well as J4 have mean radial velocities around 0 \kms. J3 shows mild to medium red-shifts for He~II, C~II, O~II, [O~III], and [Ne~IV], while He~I, [N~II], [Ar~IV], [Ar~V], and [Ne~III] are blue-shifted. For J4, He~II, C~II, N~III, O~II, [O~III], and [Ne IV] are slightly red-shifted, while He~I, [N~II], [Ar~V], [Ne~III], and [Ne IV] are blue-shifted.

\textls[-20]{The differing velocity dispersions observed in all 4 knots \mbox{(50 {\kms} $\le \sigma_{\mathrm{v_{rad}}}\le 135$ \kms)} indicate a fast stellar wind, leading to turbulence that appears to affect different ions differently. Another possible interpretation could be different kinematic components along the same line of sight.}
%The broadening of [O~III] line profile found in the overlapping region to where [Ne~IV] was detected in our data supports our hypothesis of presence of shock heating history in the region.

We have identified new recurrent structures in the velocity maps where the ions move up to $\approx$50 \kms~faster compared to the surrounding areas. The fact that {these faster-moving ions are found between the knots can be interpreted as stellar wind accelerating the low-density regions, while the high-density regions are shielded.
}%EE: please check intended meaning has been retained. - ok
All these properties point at a rather complex ionisation structure and ejection history. We strongly encourage future IFU observations with higher spectral and spatial resolution.

%Based on the varying ionisation structure and multiple kinematic components reported in this study, it strongly supports a multiple ejection history where the older ejecta: photo-ionised gas (C~II; N~III) with no recent shocks or dust formation and the recent ejecta: shocked gas ([Ne~IV], [Ar~IV] and [Ar~V]) accompanied by subsequent dust formation in the radiative cooling zones (Equatorial region). Low ionisation energy ions (C~II, He~I, [N~II], and O~II) are predominantly found in the polar knots. Ions requiring medium energies for their production ([Ne~II], He~II, [O~III] with the exception of N~III) are found to populate all 4 knots, while the ions requiring the highest energies ([Ar~IV and V], [Ne~IV]) are mainly found in the equatorial regions. Therefore, we conclude our emission line flux maps and kinematic studies hint at the presence of shock heating, overlapping shocks and photo-ionising zones, and multiple ejection events in the evolutionary history of the A30 inner knot system. 

%%%%%%%%%%%%%%%%%%%%%%%%%%%%%%%%%%%%%%
\vspace{6pt} 

\authorcontributions{Conceptualisation, Q.A.P. and K.L.C.; methodology, Q.A.P., A.R., K.L.C., K.E.; software, K.L.C., A.R.; validation, A.R.; formal analysis K.L.C., A.R., K.E.; investigation, Q.A.P., K.L.C., A.R., K.E.; resources, Q.A.P.; data curation, K.L.C., A.R.; writing---original draft preparation, K.L.C.; writing---review and editing, A.R., Q.A.P., K.L.C.; visualization, A.R., K.L.C.; supervision Q.A.P.; project administration, Q.A.P.; funding acquisition Q.A.P. All authors have read and agreed to the published version of the manuscript.}

\funding{This research was funded by Hong Kong University Grants Council (grant numbers 17326116 and 17300417). K. Exter received support from the Euro3D Research Training Network, grant number HORN-CT2002-00305.}

\dataavailability{The datasets presented in this article are not readily available because the data are part of an ongoing study. Requests to access the datasets should be directed to Q.A.P.}

\acknowledgments{We thank the anonymous referees for their insightful comments which strongly increased the scientific value of the work presented here. This work made use of Astropy\endnote{\url{http://www.astropy.org}}, a community-developed core Python package and an ecosystem of tools and resources for astronomy~\citep{astropy:2013, astropy:2018, astropy:2022}, the “Euro 3D visualisation tool”, as well as PyNeb.}
%MDPI: Notes 1 and 3 are the same. Please check and revise if necessary. - footnote 3 removed, citation moved to first mention of PyNeb; % added “Euro 3D visualisation tool”

%\acknowledgments{We acknowledge the use of software as provided by the Starlink Project which is run by CCLRC on behalf of PPARC. K. Exter acknowledges the support from the Euro3D Research Training Network, grant No.HORN-CT2002-00305. L. Christensen acknowledges the support by the German Verbundforschung associated with the ULTROS project, grant No.05AE2BAA/4.  K. Exter would like to thank Mike Barlow and Barbara Ercolano for their patience and advice, and Roger Wesson for his codes.}

\conflictsofinterest{The authors declare no conflicts of interest. The research funders had no role in the design of the study, collection, analyses, or interpretation of data, writing of the manuscript or in the decision to publish the results. %MDPI: We revised and formatted Appendix section as per layout rules, please confirm. - ok
} 
%%%%%%%%%%%%%%%%%%%%%%%%%%%%%%%%%%%%%%%%%%%%%

\appendixtitles{no} % Leave argument "no" if all appendix headings stay EMPTY (then no dot is printed after "Appendix A"). If the appendix sections contain a heading then change the argument to "yes".
\appendixstart
\appendix
\section[\appendixname~\thesection]{}
\vspace{-12pt}
%\FloatBarrier % Place before or after the figure
\begin{figure}[H]
%\centering
\includegraphics[width=0.98\textwidth]{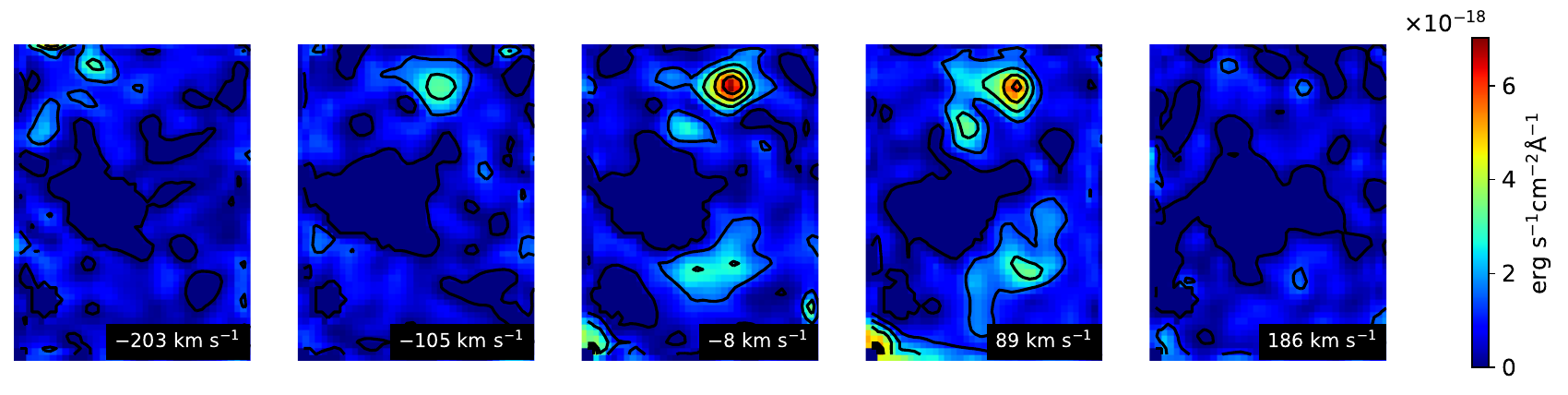}\hfill\\
\includegraphics[width=0.995\textwidth]{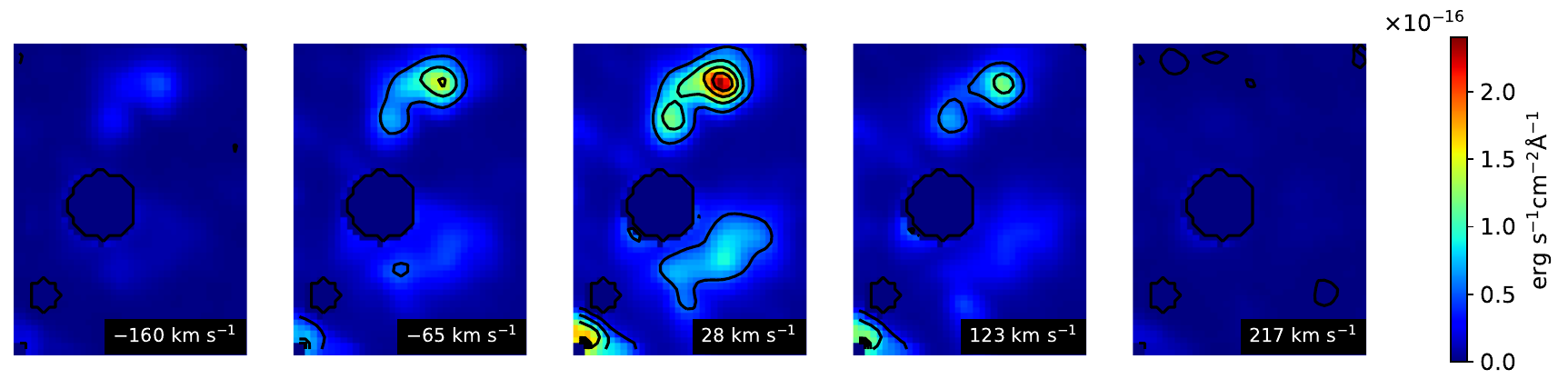}\\
\includegraphics[width=0.99\textwidth]{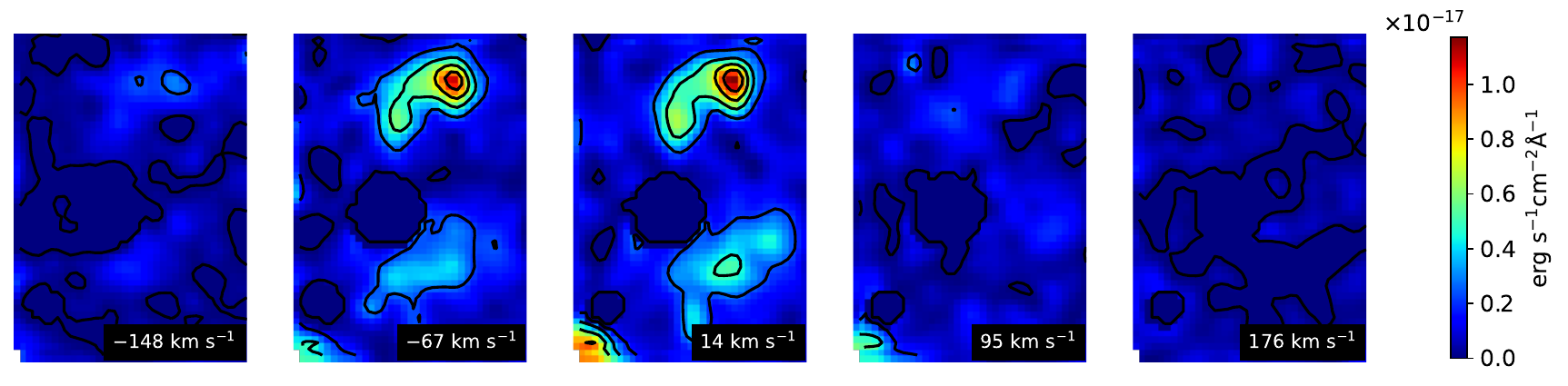}\hfill\\
\centerline{
\includegraphics[width=55mm,height=45mm]{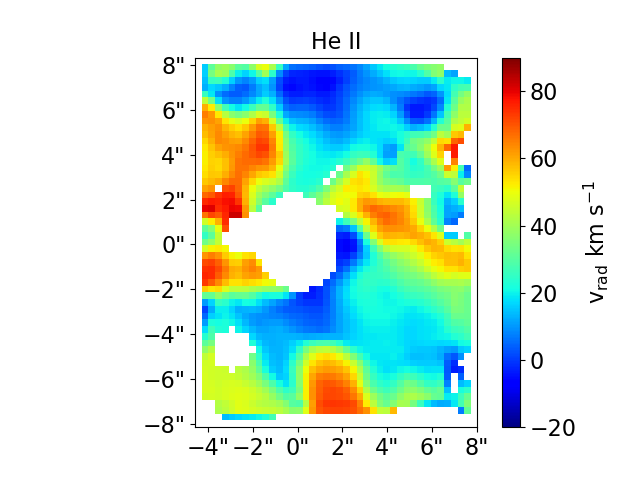}
\includegraphics[width=55mm,height=45mm]{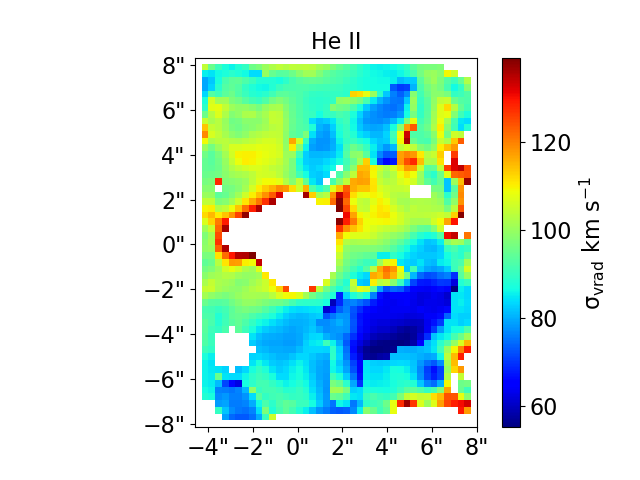}
}
\caption{Same as Figure~\ref{fig:velHeI} but for He~II (from top to bottom) 4542 \AA, 4686 \AA, and 5412 \AA.}
\label{fig:velHeII}
\end{figure}

%\FloatBarrier % Place before or after the figure
\begin{figure}[H]
%\centering
\includegraphics[width=\textwidth]{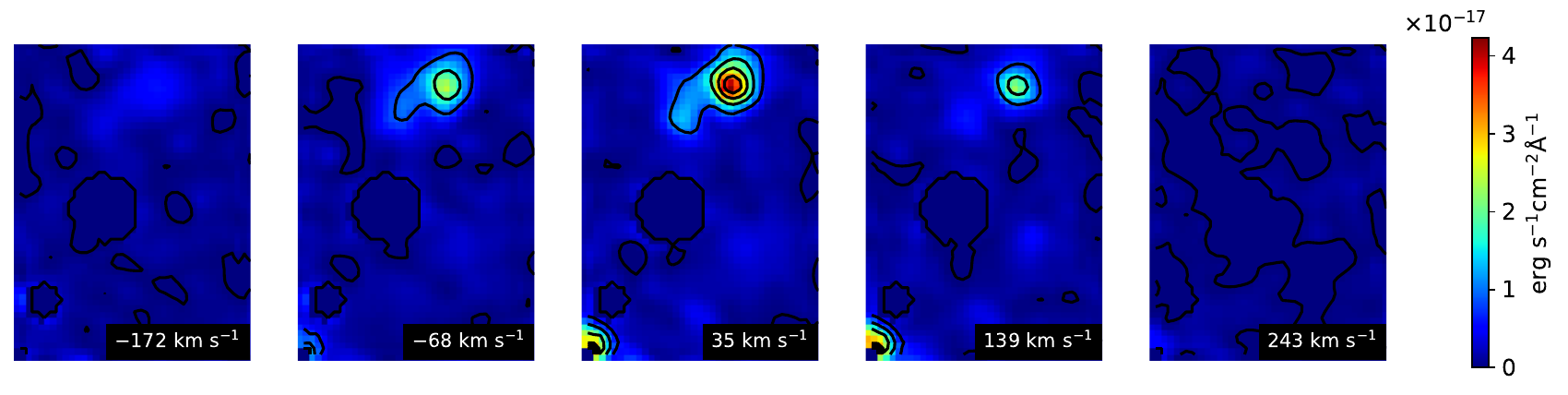}\\
\centerline{\includegraphics[width=45mm,height=45mm]{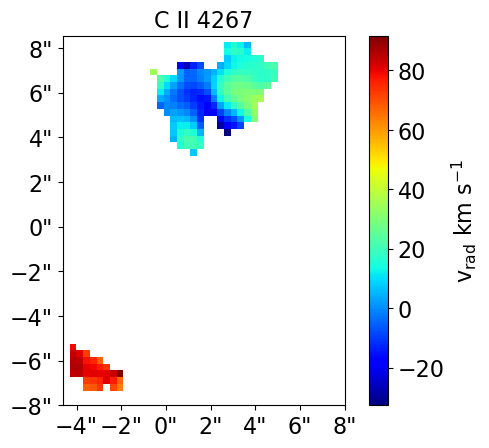}
\includegraphics[width=45mm,height=45mm]{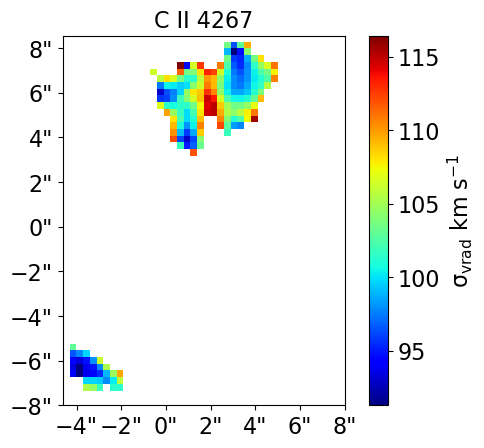}}%\hfill
\caption{Same as Figure~\ref{fig:velHeI} but for C~II 4267 \AA.}
\label{fig:velCII}
\end{figure}
\unskip
%\FloatBarrier % Place before or after the figure
\begin{figure}[H]
%\centering
\includegraphics[width=\textwidth]{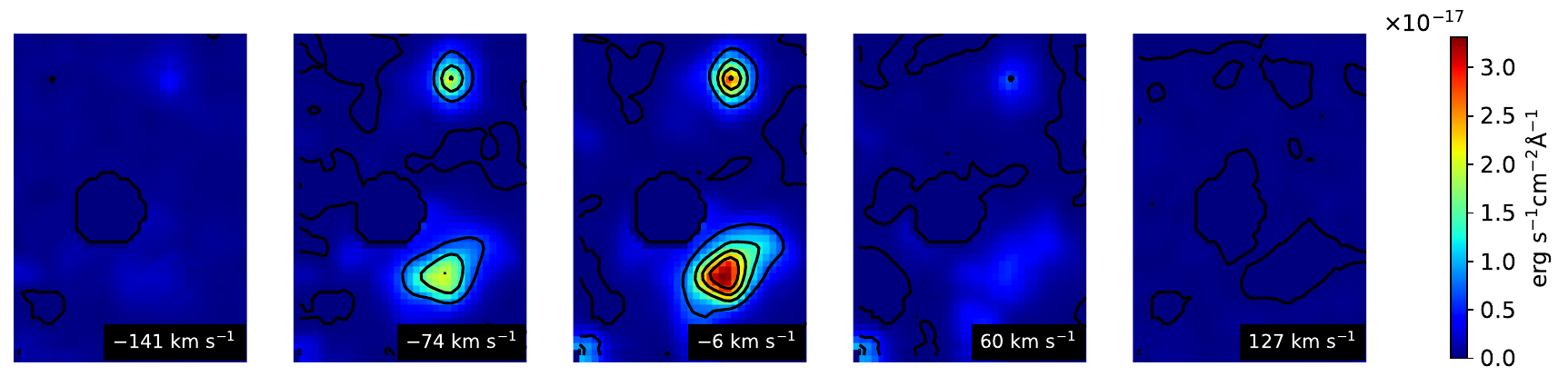}\\
\includegraphics[width=\textwidth]{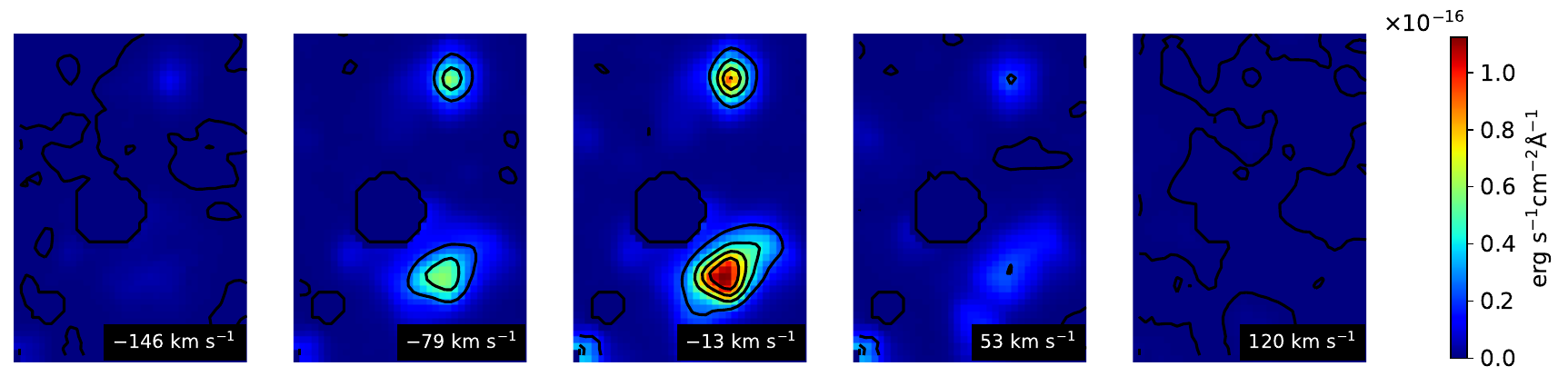}\\
\centerline{\includegraphics[width=55mm,height=45mm]{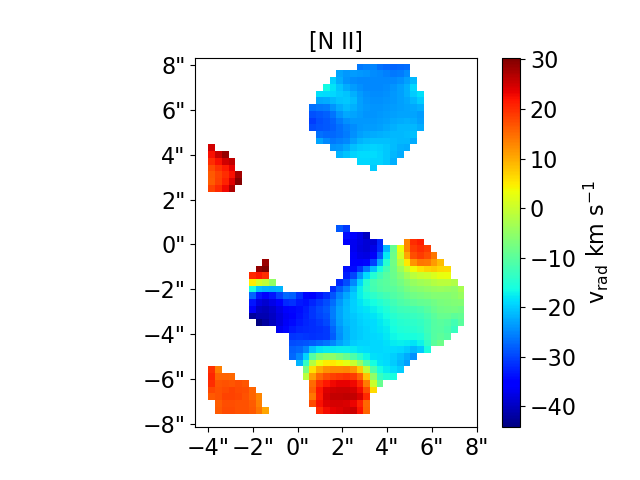}\includegraphics[width=55mm,height=45mm]{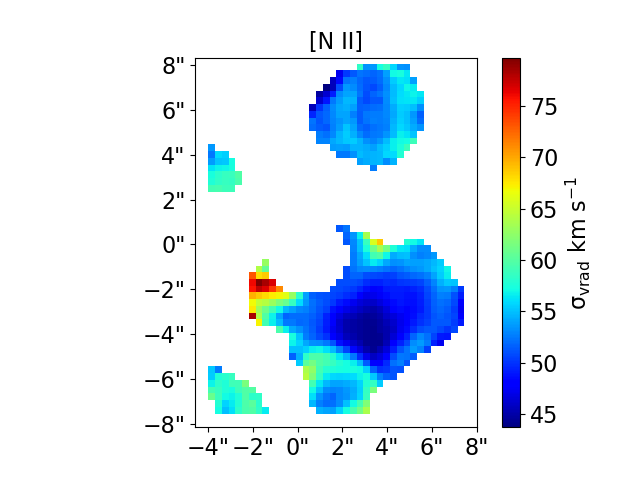}}
\caption{Same as Figure~\ref{fig:velHeI} but for [N~II] 6548 \AA~6583 \AA.}
\label{fig:velNII}
\end{figure}

%\FloatBarrier % Place before or after the figure
\begin{figure}[H]
%\centering
\includegraphics[width=\textwidth]{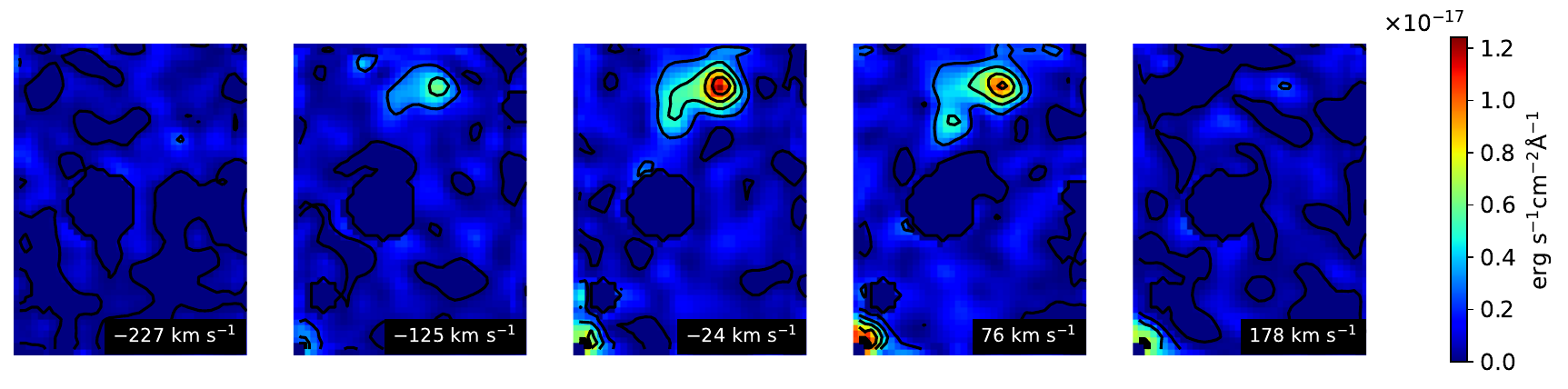}\\
\centerline{\includegraphics[width=45mm,height=45mm]{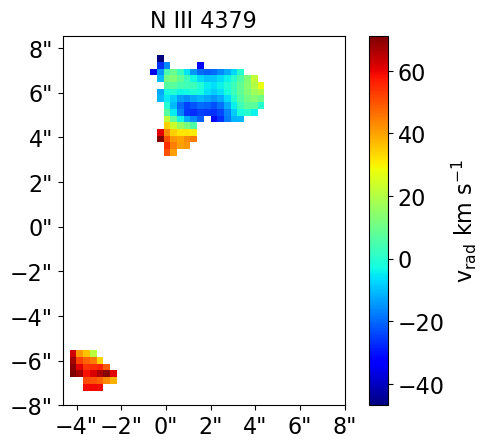}
\includegraphics[width=45mm,height=45mm]{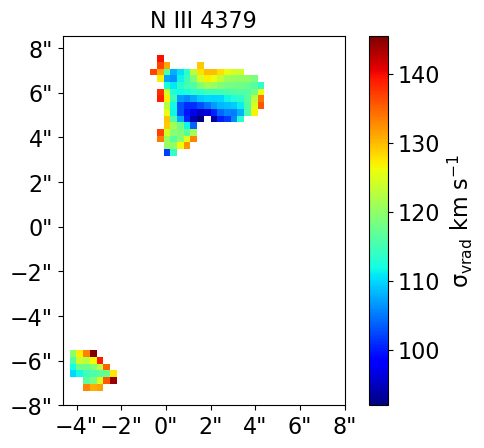}}
\caption{Same as Figure~\ref{fig:velHeI} but for N~III 4379 \AA.}
\label{fig:velNIII}
\end{figure}
\unskip
%\FloatBarrier % Place before or after the figure
\begin{figure}[H]
%\centering
\includegraphics[width=0.98\textwidth]{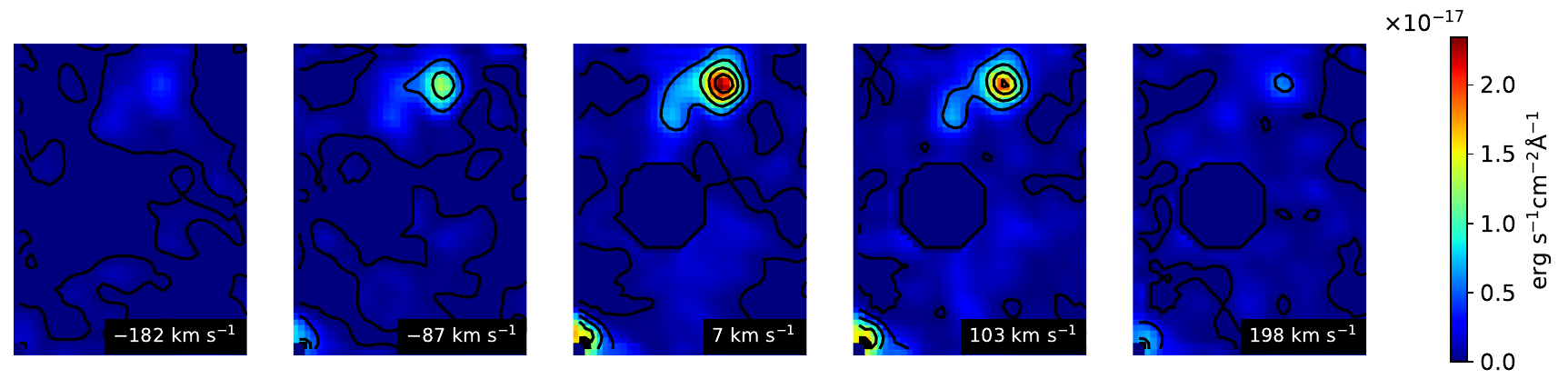}\\
\centerline{\includegraphics[width=45mm,height=45mm]{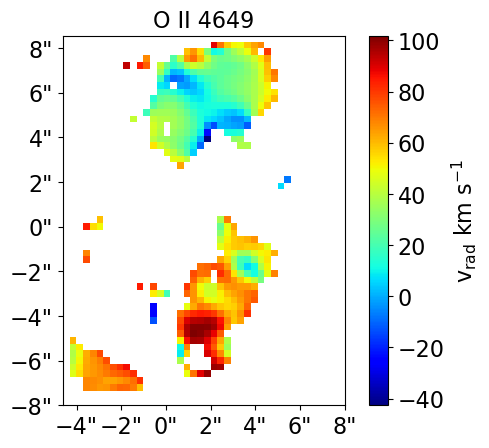}
\includegraphics[width=45mm,height=45mm]{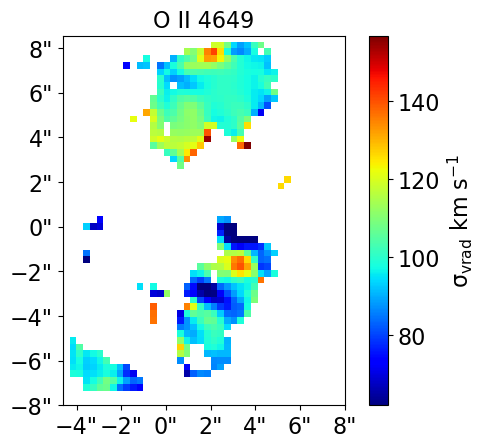}}
\caption{Same as Figure~\ref{fig:velHeI} but for O~II 4649 \AA.}
\label{fig:velOII}
\end{figure}

%\FloatBarrier % Place before or after the figure
\begin{figure}[H]
%\centering
\includegraphics[width=0.98\textwidth]{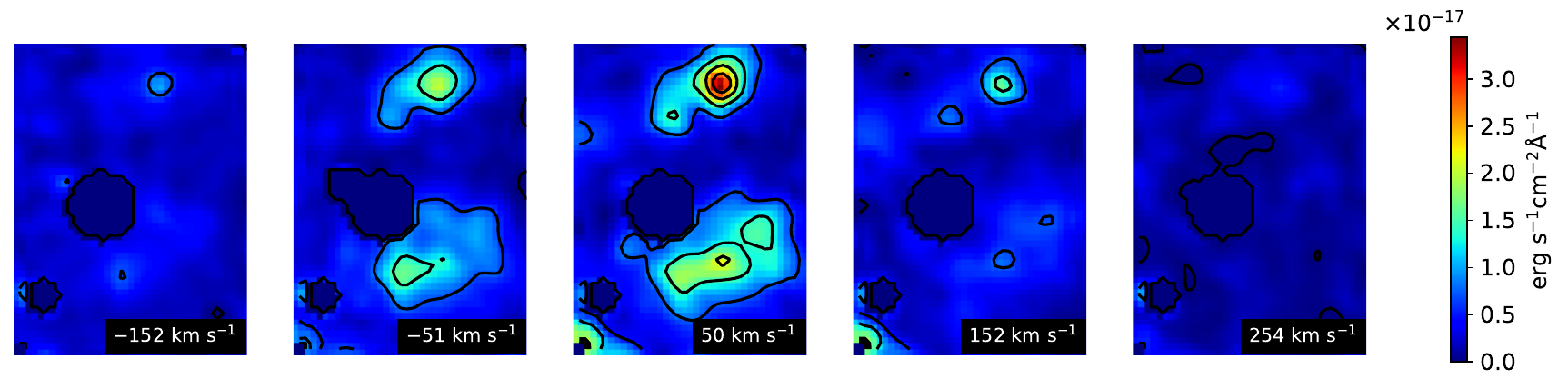}\hfill\\
\includegraphics[width=0.99\textwidth]{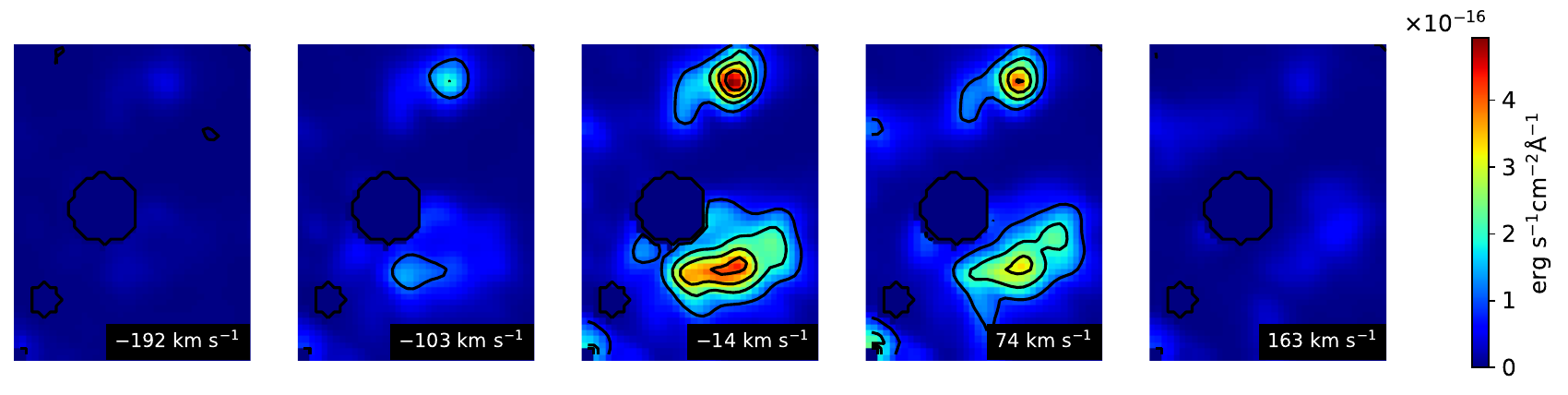}\hfill\\
\includegraphics[width=\textwidth]{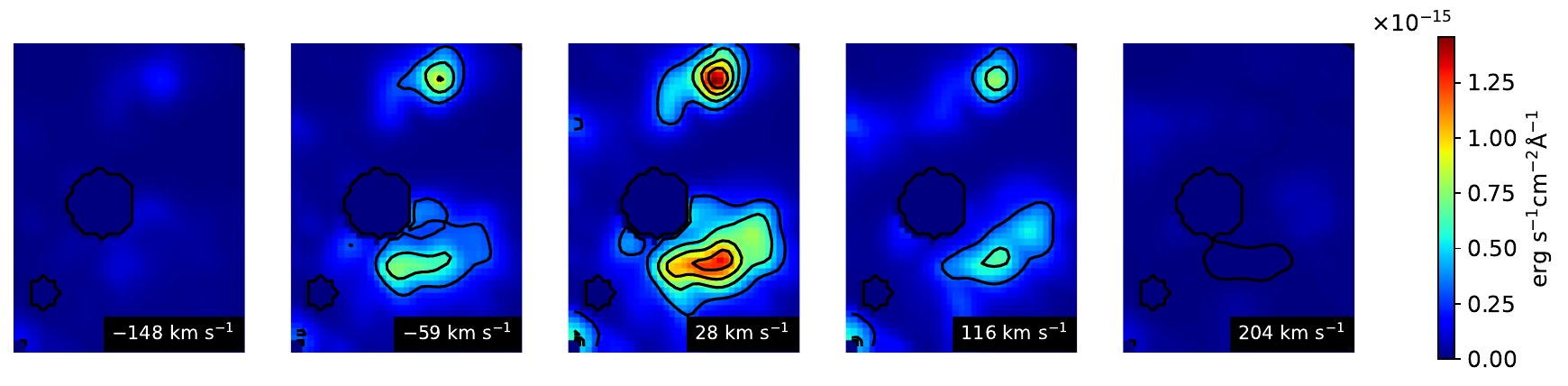}\\
\centerline{\includegraphics[width=55mm,height=45mm]{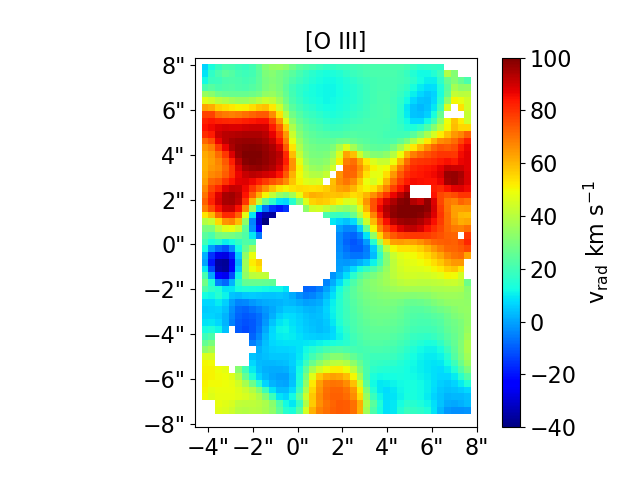}
\includegraphics[width=55mm,height=45mm]{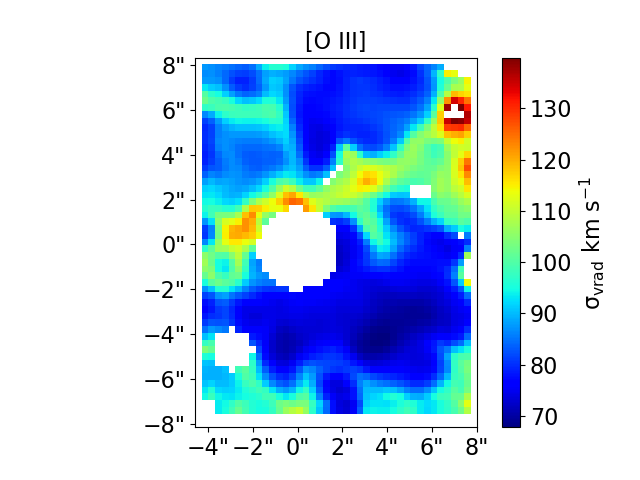}}
\caption{Same as Figure~\ref{fig:velHeI} but for [O~III] (from top to bottom) 4363 \AA, 4959 \AA, and 5007 \AA.}
\label{fig:velOIII}
\end{figure}
\unskip
%\FloatBarrier % Place before or after the figure
\begin{figure}[H]
%\centering
\includegraphics[width=\textwidth]{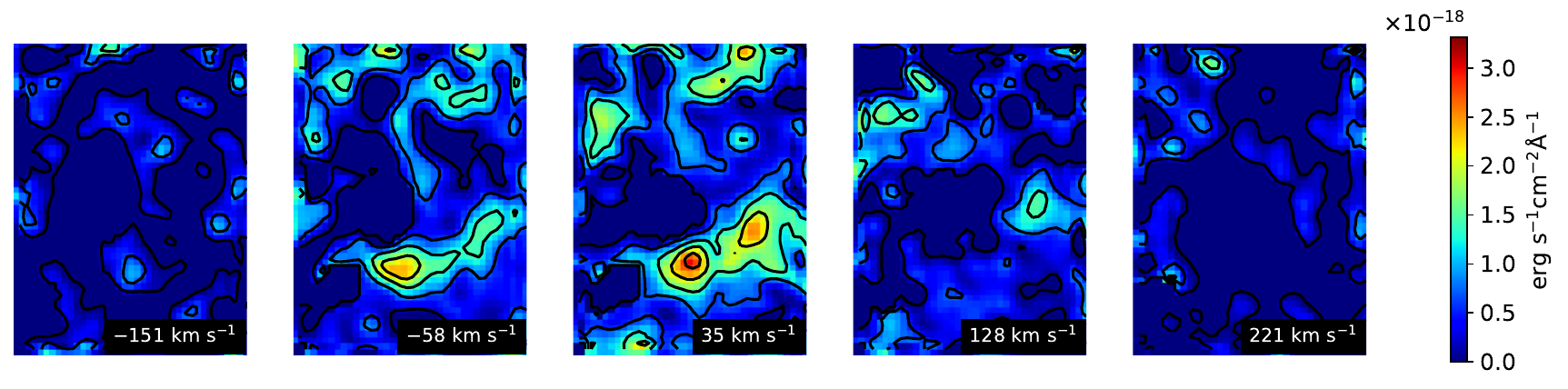}\\
\centerline{\includegraphics[width=45mm,height=45mm]{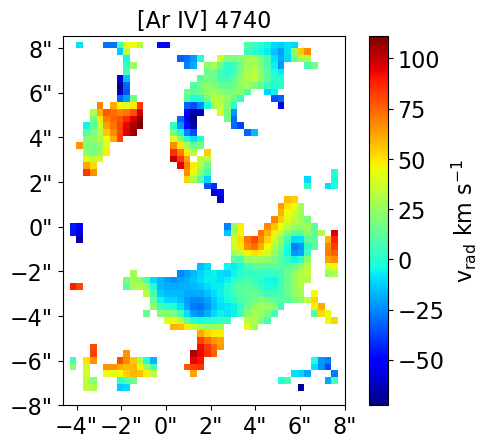}
\includegraphics[width=45mm,height=45mm]{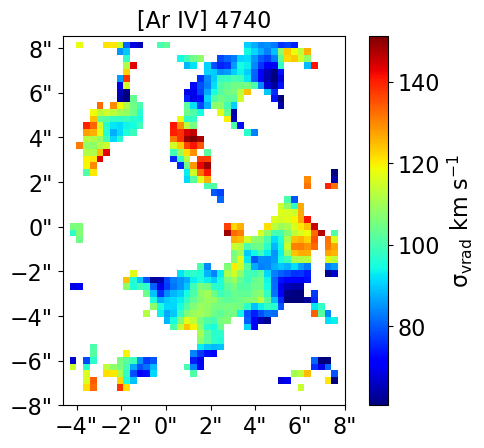}}%\hfill
\caption{Same as Figure~\ref{fig:velHeI} but for [Ar~IV] 4740 \AA.}
\label{fig:velArIV}
\end{figure}
%\FloatBarrier % Place before or after the figure
\begin{figure}[H]
%\centering
\includegraphics[width=\textwidth]{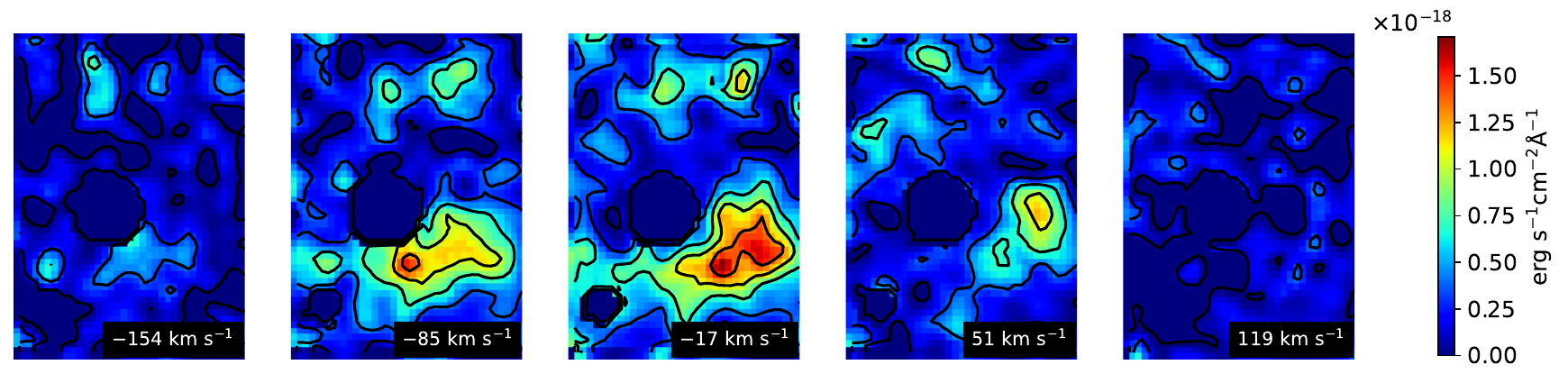}\\
\centerline{\includegraphics[width=45mm,height=45mm]{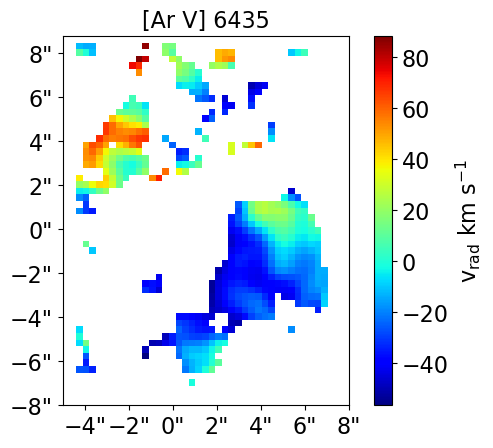}
\includegraphics[width=45mm,height=45mm]{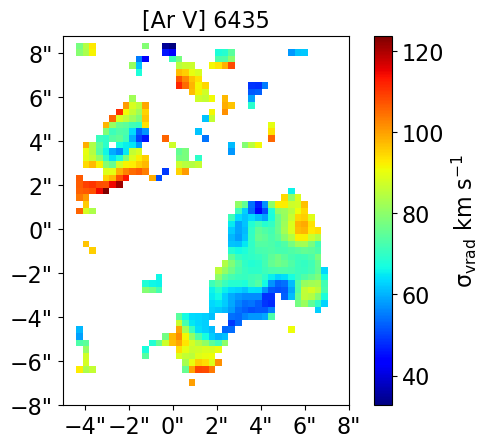}}
\caption{Same as Figure~\ref{fig:velHeI} but for [Ar~V] 6435 \AA.}
\label{fig:velArV}
\end{figure}
\unskip
%\FloatBarrier % Place before or after the figure
\begin{figure}[H]
%\centering
\includegraphics[width=\textwidth,height=35mm]{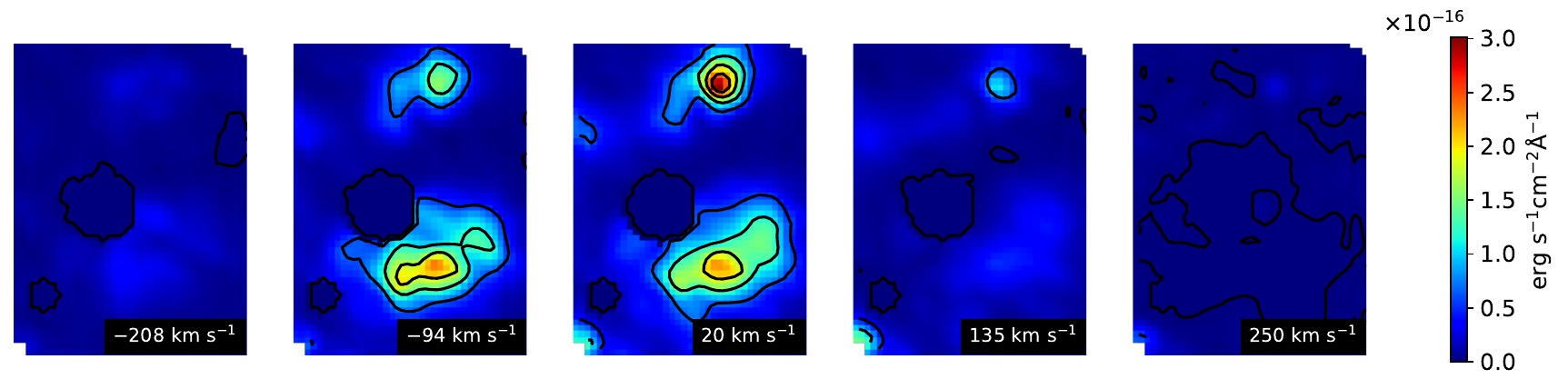}\\
\includegraphics[width=0.99\textwidth,height=35mm]{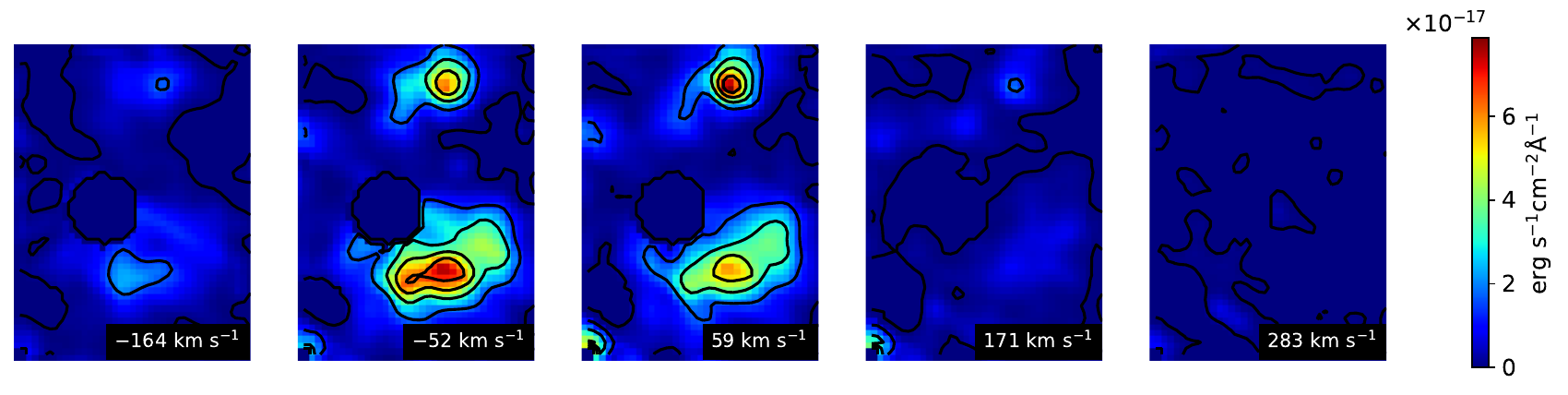}\hfill\\
\centerline{\includegraphics[width=55mm,height=45mm]{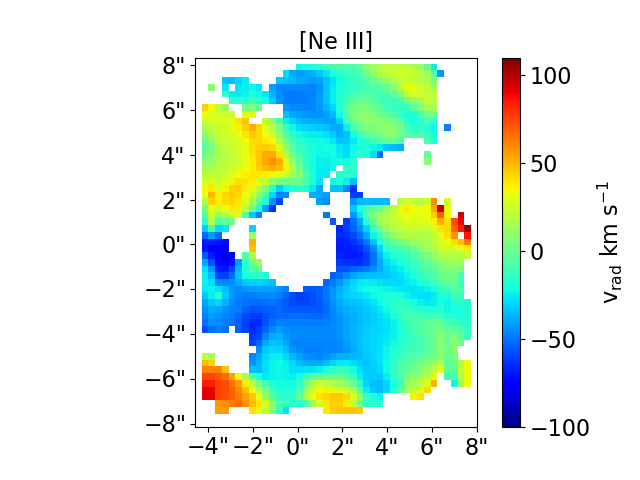}%\hfill
\includegraphics[width=55mm,height=45mm]{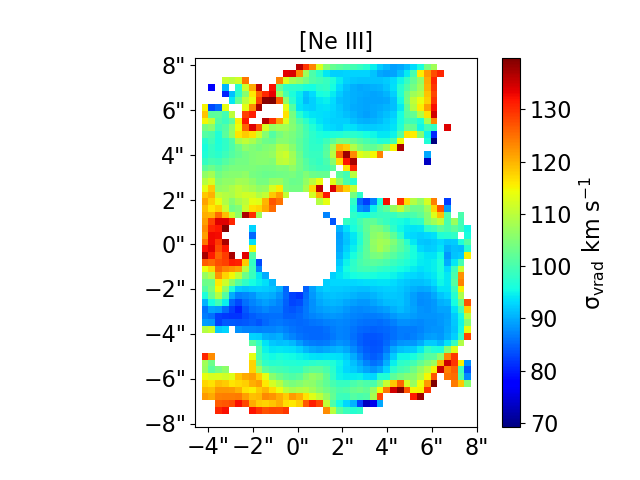}}%\hfill
\caption{Same as Figure~\ref{fig:velHeI} but for [Ne~III] 3869 \AA, 3967 \AA.}
\label{fig:velNeIII}
\end{figure}

%\FloatBarrier % Place before or after the figure
\begin{figure}[H]
\centering
\includegraphics[width=\textwidth]{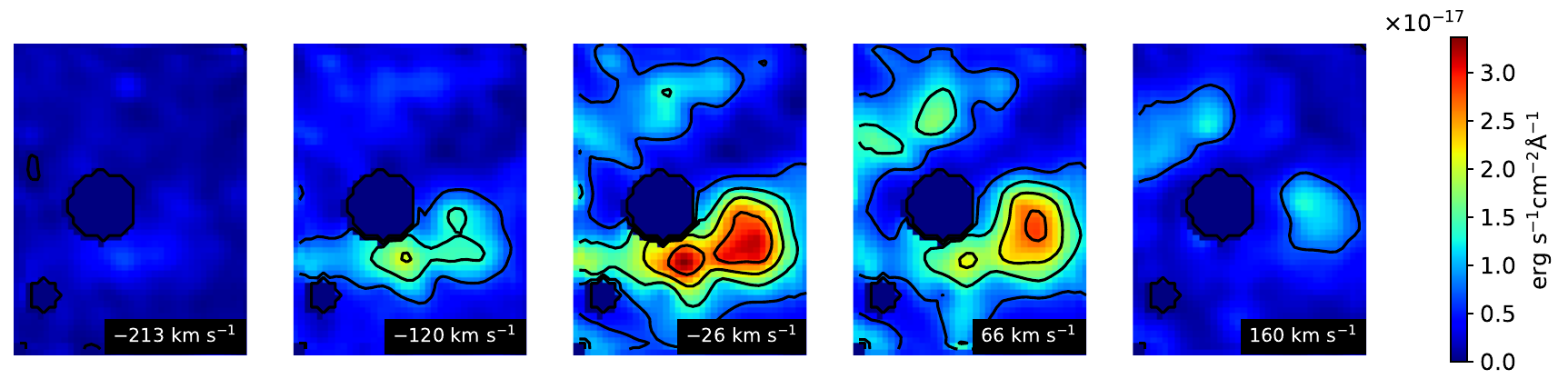}\\
\centerline{
\includegraphics[width=45mm,height=45mm]{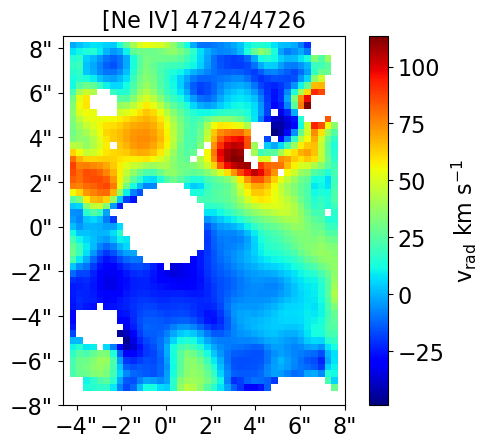}%\hfill
\includegraphics[width=43mm,height=45mm]{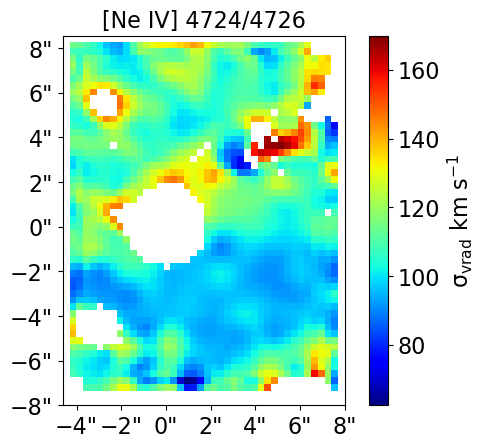}}%\hfill
\caption{Same as Figure~\ref{fig:velHeI} but for the unresolved [Ne~IV] doublet 4724/4726 \AA~(calculated relative to the assumed remaining wavelength of $\lambda$ 4724.89 \AA). Note that because of its doublet nature, the fitted line is widened, resulting in inflated velocity dispersions.}
\label{fig:velNeIV}
\end{figure}

%%%%%%%%%%%%%%%%%%%%%%%%%%%%%%%%%%%%%%%%%%
\begin{adjustwidth}{-\extralength}{0cm}
\printendnotes[custom] % Un-comment to print a list of endnotes

\reftitle{References}

\PublishersNote{}
\end{adjustwidth}
\end{document}